# Tailored Vapor Deposition Unlocks Large-Grain, Wafer-Scale Epitaxial Growth of 2D Magnetic CrCl₃

Vivek Kumar[1], Abhishek Jangid[1,#], Manas Sharma[2,#], Manvi Verma[1], Jampala Pasyanthi[2], Keerthana S Kumar[1], Piyush Sharma[2], Emil O. Chiglintsev[3,4], Mikhail I. Panin[3,4], Sudeep N. Punnathanam[2], Alexander I. Chernov[3,4], Ananth Govind Rajan[2], Akshay Singh[1,*]

[1]*Department of Physics, Indian Institute of Science, Bengaluru, Karnataka 560012, India*

[2]*Department of Chemical Engineering, Indian Institute of Science, Bengaluru, Karnataka 560012, India*

[3]*Russian Quantum Center, Skolkovo Innovation City, Moscow 121205, Russia*

[4]*Center for Photonics and 2D Materials, Moscow Institute of Physics and Technology, Dolgoprudny 141700, Russia*

[#]These authors contributed equally.

*Corresponding author: aksy@iisc.ac.in

**Abstract:**

Two-dimensional magnetic materials (2D-MM) are an exciting playground for fundamental research, and for spintronics and quantum sensing. However, their large-grain large-area synthesis using scalable vapour deposition methods is still an unsolved challenge. Here, we develop a tailored approach for centimetre-scale growth of semiconducting 2D-MM CrCl₃ films on mica substrate, via physical vapour transport deposition. A controlled synthesis protocol, enabled via innovations concerning light management, very-high carrier-gas flow, precursor flux, and oxygen/moisture removal, is critical for wafer-scale growth. Optical, stoichiometric, structural, and magnetic characterization identify crystalline, phase-pure 2D-MM CrCl₃. Substrate temperature tunes thickness of films from few-layers to tens of nanometres. Further, selective-area growth and large-area transfer are demonstrated. Substrate-dependent growth features are explained by density functional theory and state-of-the-art machine learning interatomic potential-based atomic-scale simulations. This scalable vapour deposition approach can be applied for growth of several 2D-MM, and low growth temperature (~500 °C) will enable creation of hybrid heterostructures.

**Introduction:**

Two-dimensional magnetic materials (2D-MM) exhibit intrinsic long-range magnetic order, first discovered in $Cr_2Ge_2Te_6$ and $CrI_3$[1,2]. van der Waals (vdW) 2D-MM demonstrate remarkable properties, such as strain and doping-induced magnetic phase transitions and layer-dependent magnetic ordering[3–5]. For example, $CrCl_3$ has in-plane ferromagnetic ordering but inter-layer antiferromagnetic interactions[6,7]. Heterostructures of vdW 2D-MM with other 2D materials exhibit interfacial phenomena such as exchange bias and proximity effects, and lead to spintronic and computing applications[8–10]. Interestingly, semiconducting 2D-MM enable gate-controlled spin field-effect transistors and sensors, and can be integrated with standard electronic processes[11].

Wafer-scale growth of 2D-MM is critical for hetero-integration and for scaling towards applications. Further, epitaxially-aligned large grain sizes are needed for intrinsic magnetic properties and coalesced films, not constrained by large-angle grain boundaries[12]. Mechanically exfoliated micrometre-sized flakes suffer from lack of layer-number control, and scalability and reproducibility issues. Conversely, large-scale synthesis of 2D-MM is challenging due to their low stability in ambient conditions, leading to exponentially harder synthesis at high temperatures even with trace presence of oxygen or moisture. Few studies using molecular beam epitaxy (MBE) have grown chalcogenide-based, and mostly metallic 2D-MM, but are limited by intrinsic issues with MBE: only few precursors are permissible (limited to no halide precursors), small grain sizes (~ 30 nm), and slow growth rates[13,14]. In contrast, physical vapor transport deposition (PVTD) and chemical vapor deposition (CVD) studies have reported small-scale isolated $CrCl_3$ and $CrSe_2$ flakes, respectively[15,16]. Existing studies on 2D materials demonstrate the critical role of substrate in facilitating growth and vdW rotational epitaxy, where atomic structure, crystalline nature and topology of substrate vs overlayers critically influence growth[17–19]. Such insights for 2D-MM are critically missing, and can lead to tuning of magnetic ordering and increased functionality via vapour deposition processes[3,19].

We report the centimetre-scale growth of semiconducting 2D-MM $CrCl_3$ via a tailored PVTD process. $CrCl_3$ is grown on several substrates, with large-grain (~10 $\mu$m), epitaxial, and coalesced films grown on fluorophlogopite-mica (F-mica: $(KMg_3(Si_3Al)O_{10}F_2)$). The thickness (layer-number) increases with temperature, whereas flow rate of carrier gas controls nucleation density and coverage. This remarkable result is enabled by synthesis innovations, specifically the control of light, oxygen and moisture, and precursor flux in the growth tube. Optical spectroscopy (Raman, photoluminescence (PL)) confirm the grown flakes as $CrCl_3$, and SEM-EDS (scanning electron microscopy-energy dispersive x-ray spectroscopy) and X-ray photoelectron spectroscopy (XPS) confirm the stoichiometry. Scanning transmission electron microscopy (STEM) demonstrates high structural order and monoclinic single-crystallinity. SQUID (superconducting quantum interference device) measurements match the expected magnetic ordering. Additionally, density functional theory (DFT) calculations and machine learning molecular dynamics (MLMD) simulations based on machine-learned interatomic potentials (MLIPs) enable deep understanding of substrate-dependent growth. The foundation MLIPs are recently released state-of-the-art models, which are trained on millions of data points, and enable large-scale MD simulations of PVTD precursor dynamics with first-principles accuracy[20–22]. Finally, both transfer of grown films, and patterned growth are demonstrated, enabling device applications. The low deposition temperature (~500 °C) will enable ready integration with other 2D materials for creating functional heterostructures[23].

**Growth of 2D-MM films and characterization**

$CrCl_3$ powder precursor is placed in the first zone (650-750 °C), and substrates are kept in the second zone (450-600 °C) of the furnace (Fig. 1a, Methods and Supplementary Section 1 for growth details). Given the sensitivity of $CrCl_3$ to oxygen and moisture, inert carrier gas (argon) is passed through a two-stage trace moisture and oxygen filter, followed by complex flushing and purging processes. The tubing connections follow high-vacuum compatibility protocols (Methods, Supplementary Section 2). Additionally, $CrCl_3$ and other 2D-MM undergo redox reactions in presence of light and moisture/oxygen, and hence reactor is kept in a dark fumehood[24,25]. Notably, aluminium foil wrapped around outside of the growth tube in the second zone, to prevent light from heating elements to reach the substrate, has striking benefits on growth (discussed later). Resulting from this extensive optimization, centimetre-scale growth of $CrCl_3$ is achieved (Fig. 1b-c), with easily observable contrast differences of F-mica substrate, before and after growth (Fig. 1b). Contrast is homogeneous, indicating uniform large-scale growth (Supplementary Section 3 shows optical microscope (OM) images).

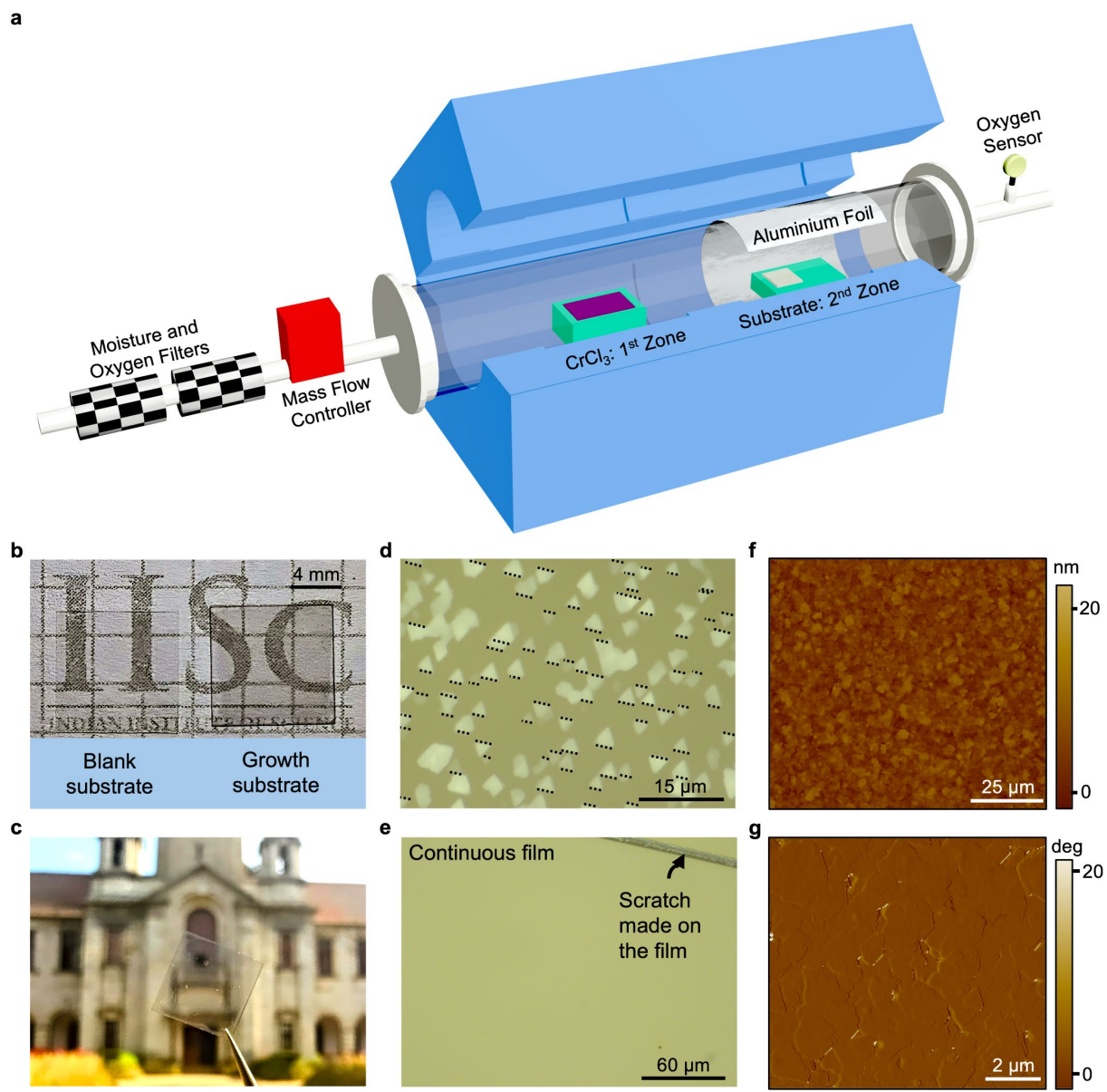

**Fig. 1: Epitaxial large-scale growth of 2D-MM. a,** Two-zone physical vapor transport deposition (PVTD) setup. **b,c,** Wafer-scale growth of two-dimensional magnetic materials (2D-MM) $CrCl_3$. Indian Institute of Science's main building is shown in the background of **(c)**. **d,e,** Optical microscope (OM) images of typical optimized growth observed on fluorophlogopite mica (F-mica) substrate. **d,** Thin, directionally aligned epitaxial growth. **e,** Continuous coverage of $CrCl_3$ film. **f,** Atomic force microscope (AFM) image for the large-scale 2D-MM growth. **g,** AFM phase map image acquired for the large-scale growth shows minimal grain boundaries.

To visualize the grown flakes on F-mica, we show a short-time growth. The flakes are triangular, and black dotted lines are drawn at the edge (Fig. 1d). The lines are parallel to each other, indicating rotational-epitaxy of CrCl$_3$ on F-mica. With further optimization, epitaxially grown flakes converge to form a continuous film (Fig. 1e). To the best of our knowledge, this is the first demonstration of large-scale epitaxial growth of any vdW 2D-MM using scalable vapour deposition methods. Atomic force microscopy (AFM) scans on the continuous films (Fig. 1f) show reasonably smooth topography (RMS roughness ~ 2 nm), desirable for creating heterostructures with other 2D materials (Methods, and Supplementary Section 4 for AFM). Minimal grain boundaries are observed in AFM phase mapping (Fig. 1g) due to epitaxial growth.

We next investigated the stoichiometry, crystallinity, and structure of grown 2D-MM. CrCl$_3$ exists in monoclinic (C2/m) phase at room temperature (Fig. 2a)[26,27]. Optical spectroscopy (Raman, PL) is performed in a custom vacuum chamber to minimize laser-induced sample damage. Raman spectra (Fig. 2b) are taken on transferred flakes on SiO$_2$/Si to avoid the strong F-mica background (Methods), and shows Raman peaks consistent with monoclinic CrCl$_3$[15]. PL spectra (Fig. 2c) on the as-grown film shows a broad peak around 1.4 eV, arising from $^4T_2$ to $^4A_2$ d-d transition in CrCl$_3$[6,28]. SEM-EDS (Fig. 2d) shows a near-ideal stoichiometric ratio of ~ 0.34 between Cr and Cl. XPS performed on as-grown CrCl$_3$ film shows minimal oxidation-related peaks around the Cl-2p peak (Fig. 2e, Supplementary Section 5). The Cl-2p spectrum exhibits a spin-orbit doublet at 199.19 eV (Cl-2p$_{3/2}$) and 200.80 eV (Cl-2p$_{1/2}$), with the splitting consistent with pristine CrCl$_3$[29]. Minor contributions from Cl vacancies (197.95 eV) and surface O-(CrCl$_3$) (200.12 eV) indicate minimal defects/oxidation. Cr-2p analysis is limited by multiplet complexity, but aligns with CrCl$_3$ reference data (Supplementary Section 5)[30].

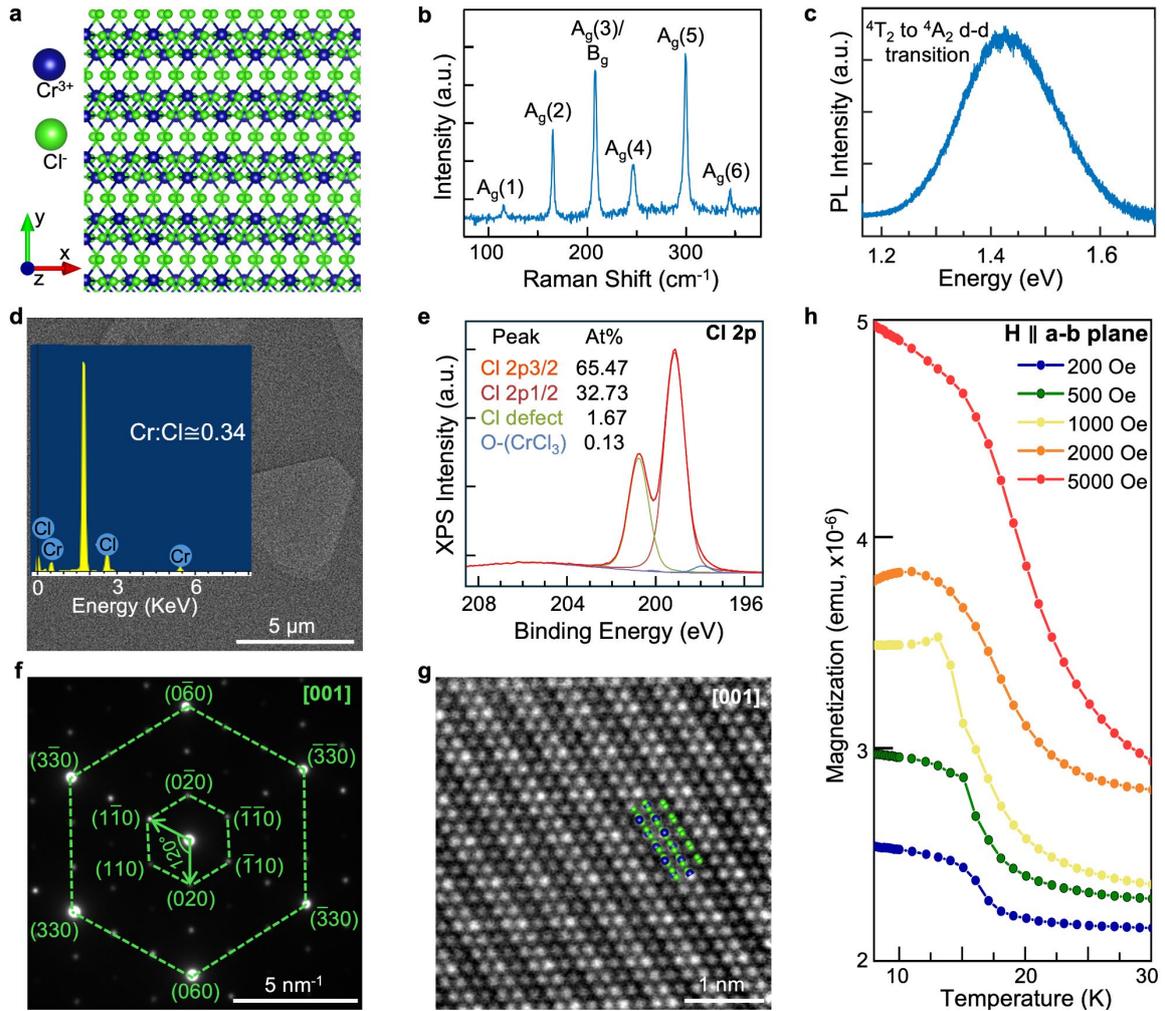

**Fig. 2: Stoichiometry, crystallinity, and magnetic characterizations of the grown CrCl$_3$ films. a,** Top view schematic of three layers of CrCl$_3$, where the monolayer is a honeycomb lattice- with one Cr atom connected to six Cl atoms. **b,** Raman spectra of a CrCl$_3$ flake transferred onto SiO$_2$/Si, with characteristic CrCl$_3$ Raman peaks. **c,** Photoluminescence (PL) spectra shows a broad peak around 1.4 eV, characteristic of CrCl$_3$. **d,** Scanning electron microscopy energy dispersive X-ray spectroscopy (SEM-EDS) spectra show a near-perfect stoichiometry of ~1:3 (Cr:Cl). Inset shows the corresponding SEM image. **e,** X-ray photoelectron spectra (XPS) of Cl-2p peak shows minimal evidence of oxidation. **f,** Selected area electron diffraction (SAED) micrograph matches expected pattern for CrCl$_3$. **g,** High-angle annular dark field-scanning tunnelling electron microscopy (HAADF-STEM) image matches with a monoclinic lattice and shows the expected periodicity. Inset shows an overlay of atomic structure. **h,** Magnetization versus temperature for in-plane fields of 200-5000 Oe. The curves for (500, 1000, 2000, and 5000 Oe) are offset in the y-direction by (0.13, 0.2, 0.6, and 0.66) × 10$^{-6}$ emu, respectively.

For TEM, $CrCl_3$ flakes are transferred to a holey SiN grid (Methods). SAED pattern (Fig. 2f) matches the reported and simulated pattern for monoclinic $CrCl_3$ (Supplementary Section 6)[15]. Sharp SAED spots corresponding to $CrCl_3$, and no signature of impurity phases, reinforce the single-crystalline nature of synthesized flakes. HAADF-STEM (Fig. 2g) shows a centred honeycomb lattice that matches with the simulated image for monoclinic $CrCl_3$, with a near-ideal inter-planar d-spacing of 5.13 Å (Supplementary Section 6)[15].

To investigate magnetic properties, SQUID magnetometry was performed on $CrCl_3$ flakes transferred on a GaAs substrate (pure diamagnetic, see Methods). We note that magnetic properties of thin flakes may vary c.f. bulk crystals due to coexistence of monoclinic and rhombohedral phases at low temperatures[28]. Fig. 2h shows the magnetic moment as a function of temperature for in-plane fields of 200-5000 Oe for zero-field cooling (diamagnetic contribution of GaAs was subtracted). The expected sharp kink in magnetization near 14 K, indicating antiferromagnetic-to-ferromagnetic transition, is present for both in-plane and out-of-plane magnetic field configurations (Supplementary Section 7). For in-plane orientation, the kink disappears at a strong magnetic field (5000 Oe), and for out-of-plane orientation, the kink becomes wide and small in weak fields (200-500 Oe), matching reports on exfoliated $CrCl_3$ (M-H curves at different temperatures in Supplementary Section 7 have similar trend)[31]. These magnetic and structural characterizations validate our growth technique for high-quality $CrCl_3$, not limited by oxidation[32].

**Innovations enabling the growth of large-area 2D-MM**

At high temperatures, stability of 2D-MM reduces in presence of light (discussed earlier). As shown in Fig. 3a (left panel), the as-grown film displays clear signs of etching, where the only source of light is radiative flux from heating elements of furnace. Careful choice of a heater material with low emissivity, but high thermal conductivity, can help in reducing radiative versus conductive heating. We choose aluminium foil, with low emissivity ~ 0.2 (c.f. 0.7 for heating elements), and wrap it around the outside of reactor tube (Fig. 1a). Fig. 3a (right panel) shows dramatic reduction of etching in growths performed with this low-emissivity secondary heating source.

Lateral size and thickness of $CrCl_3$ flakes increases with increasing substrate temperature (Fig. 3b and Supplementary Section 8), similar to other 2D-MM growth[33]. Lateral size increase is attributed to enhanced migration of precursors over substrate at higher temperatures[34,35]. Thickness increase is attributed to lower surface free energy of $CrCl_3$ c.f. substrate at higher temperatures, thus favouring vertical growth. Temperature tuning allows growth from few-layers to tens of nanometres.

The growth improves dramatically at very high flow rates (~ 500 sccm), leading to nearly coalesced films in very short times ~ 5 minutes (Fig. 3c and Supplementary Section 8). This unexpected observation indicates the sensitive balance between kinetic and thermodynamic processes. With increased flow rate and precursor supply, nucleation density and coverage increase, with a corresponding decrease in lateral size of flakes. Meanwhile, at a shorter growth time, thinner but slightly unfaceted growth is observed. Interestingly, lateral size of flakes remains nearly constant with growth time. These observations are characteristic of diffusion-limited growth, reaching saturation early in the process, limited by temperature-controlled migration barrier of $CrCl_3$ on substrate (supported by our atomic-scale simulations, see below)[36]. Increase in thickness with growth time is attributed to vertical growth of $CrCl_3$, after lateral growth saturates, under continuous precursor supply.

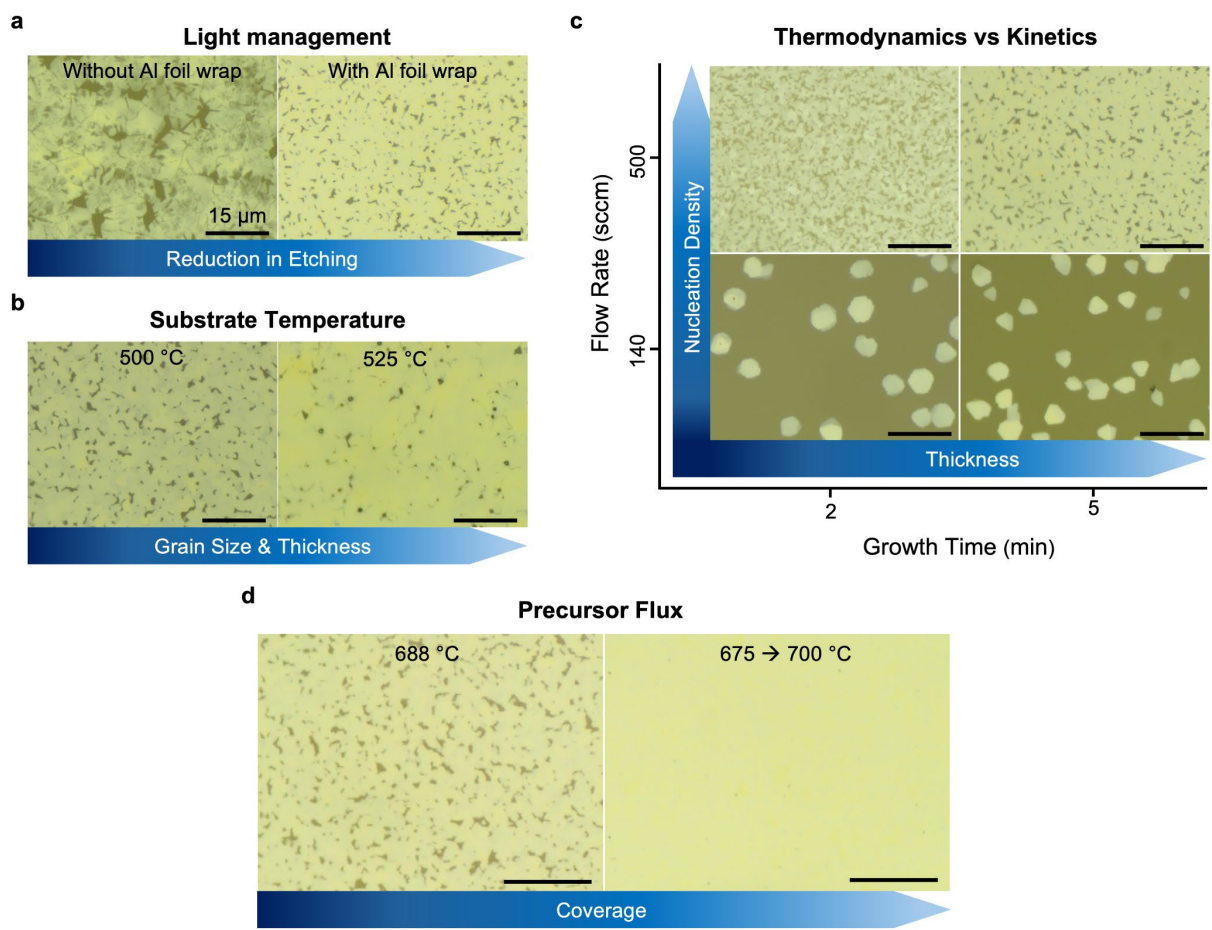

**Fig. 3: Effect of growth conditions on the observed growth of 2D-MM on F-mica substrate. a,** Left (right) panel shows an OM image of a typical growth without (with) aluminium foil wrapping around the growth tube and shows etching (no etching). **b,** The effect of substrate temperature. Lateral size and thickness of grown $CrCl_3$ flakes increase with increasing substrate temperature. **c,** The effect of flow rate and growth time. The nucleation density (number of flakes per unit area) and coverage increase with increasing flow rate. The size of grown flakes remains nearly constant, but thickness increases with increase in growth time. **d,** Effect of precursor temperature. Left (right) OM image corresponds to growth with fixed (gradually increasing) precursor evaporation temperature, showing discontinuous (continuous) $CrCl_3$ film. All scale bars are 15 $\mu$m.

To grow uniform and continuous films, going beyond the diffusion-limited saturation regime is required. At a fixed concentration (Fig. 3d (left panel)), coverage of the growing films saturates, resulting in a discontinuous film. However, with a gradually increasing precursor concentration (Fig. 3d (right panel)), a continuous film covering the entire substrate is grown. Gradually increasing precursor temperature increases precursor flux, resulting in increased nucleation and precursor diffusion from boundary layer to substrate, thus progressing beyond diffusion-limited growth[37,38]. Thus, collectively, the above advances lead to large-area growth of $CrCl_3$ (Supplementary Section 9).

**Understanding the Epitaxial Growth**

To understand the observed epitaxy, we chose three substrates: 1) amorphous $SiO_2$/Si, 2) crystalline sapphire, and 3) vdW crystalline mica (Supplementary Sections 8 and 10). Typical growth on $SiO_2$/Si substrate shows distribution of thick, faceted 2D $CrCl_3$ flakes, in addition to 3D deposition (Fig. 4a). On sapphire, majority of $CrCl_3$ flakes grow horizontally, are thinner and well-faceted, apart from some 3D depositions (Fig. 4b). Best growth is achieved on F-mica, with epitaxial and large triangular grains (Fig. 4c).

To provide a comprehensive understanding of observed substrate-dependent growth features, we employed a combination of atomic-scale simulation techniques, specifically DFT and MLMD simulations based on cutting-edge MLIPs. First, DFT is used to calculate adsorption energy of a $CrCl_3$ molecule on different crystalline substrates. Adsorption energies show that $CrCl_3$ binds more strongly to sapphire (−5.768 eV) than to mica (−0.455 eV), as also indicated by the shorter $CrCl_3$–sapphire distance (Fig. 4d vs 4e), thus resulting in reduced surface mobility. Weaker binding on mica allows greater molecular movement, favouring uniform film formation and epitaxial growth.

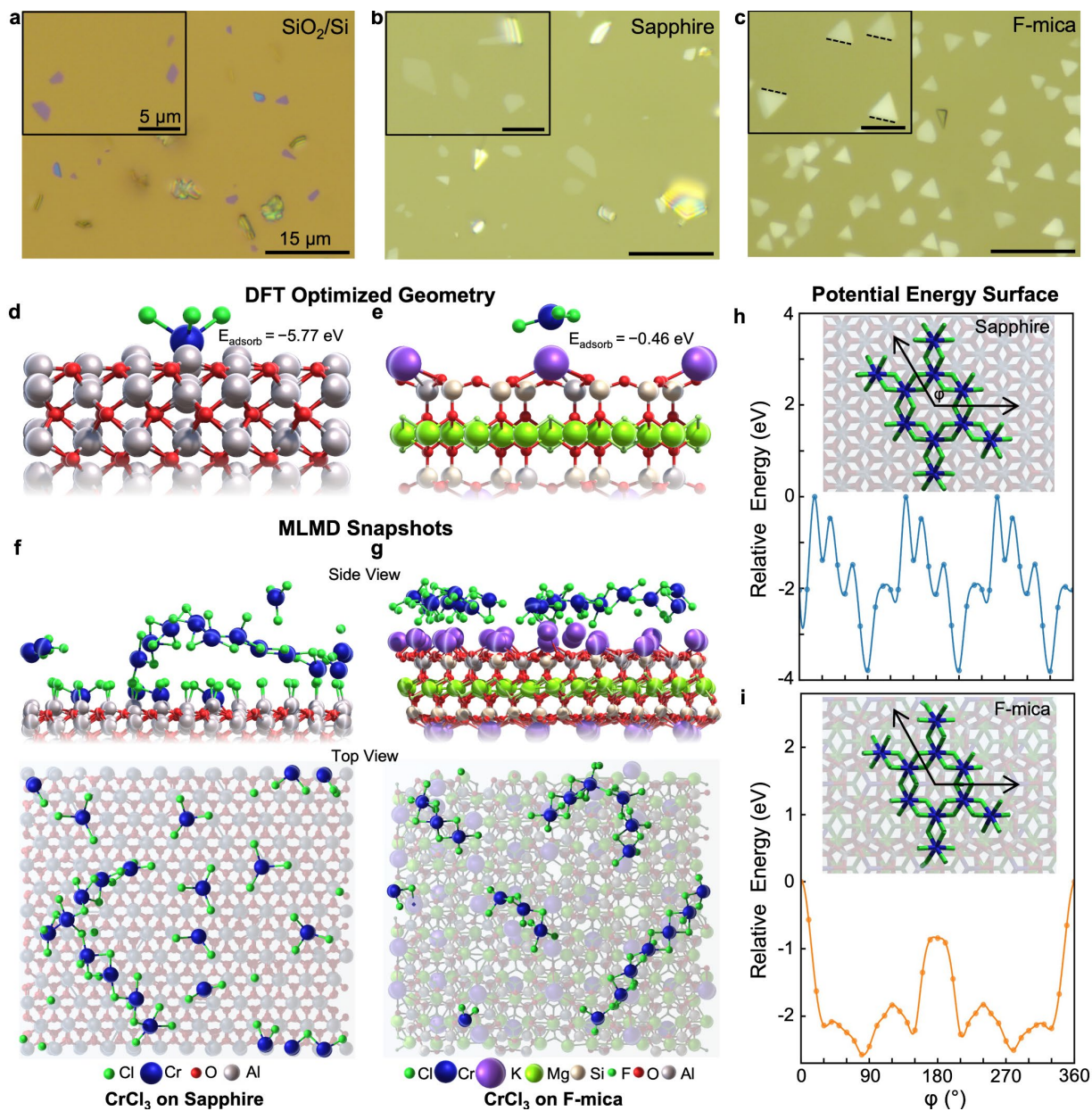

**Fig. 4: Growth on different substrates. a,** Relatively thick 2D CrCl$_3$ flakes, in addition to 3D depositions, are observed on SiO$_2$/Si. **b,** Well-faceted thinner flakes, except for a few 3D depositions, are observed on the Al$_2$O$_3$ substrate. **c,** Thin, 2D, directionally-aligned flakes are grown on the F-mica substrate. a-c, all scale bars are 15 $\mu$m. **d,e,** DFT-optimized adsorption geometries on **d,** sapphire, **e,** F-mica. **f,g,** Side and top-views of MLMD snapshots at 40 ps for 20 CrCl$_3$ molecules on **f,** sapphire, **g,** F-mica. **h,i,** Potential energy surface (PES). Relative potential energy vs. in-plane rotation angle ($\varphi$ - shown in inset) for a CrCl$_3$ layer on **h,** sapphire, **i,** F-mica from MLIP calculations.

To understand the collective behaviour of multiple CrCl$_3$ molecules, larger-scale atomistic simulations (> 1000 atoms) are necessary. In this situation, DFT or ab-initio MD (AIMD) are computationally prohibitive, and thus, foundation machine learning models are utilised[20–22]. After benchmarking and validation against DFT calculations (Supplementary Section 11), the state-of-the-art multi-atomic cluster expansion (MACE) OMAT-0 MLIP model is selected[20]. MLMD simulations with 20 CrCl$_3$ molecules are performed on sapphire and F-mica in the *NVT* ensemble at 800 K. On sapphire, molecules exhibit limited mobility, with some remaining fixed even after 40 picoseconds of simulation. Clustering into elongated, arch-like chains is observed, while isolated molecules stayed close to the surface (Fig. 4f; Supplementary Section 12), explaining the mixed 2D/3D growth morphology. Additionally, some chlorine atoms were observed to detach and stick to the substrate. On F-mica, nearly all molecules cluster and move freely across the surface (Fig. 4g), consistent with low adsorption energy. These clusters remain at ~3.2 Å from the surface, aligning with experimentally observed planar films (MLMD movies in Supplementary Material). Further, potential energy surface (PES) for lateral translation of a CrCl$_3$ molecule reveals high barriers on sapphire (~10 eV) and substantially lower ones on F-mica (~0.05 eV) (Supplementary Section 13).

To understand the epitaxially-aligned growth on F-mica, PES is computed using MLIP for a large Cr$_{10}$Cl$_{40}$ cluster (with Cr-coordination same as bulk structure), rotated about the axis normal to substrate surface. PES on sapphire (Fig. 4h) reflects the three-fold symmetry of sapphire, with deep valleys at 90º, 210º, and 330º along with numerous wells in-between. Multiple local minima, coupled with limited molecular mobility revealed above, can trap clusters in sub-optimal orientations, hindering the rearrangement necessary for epitaxial alignment. In contrast, PES on F-mica (Fig. 4i) exhibits mirror symmetry about the ac-plane, with two symmetric global minima (80º and 280º), separated from the local minima by shallow barriers, enabling facile rotation. This Goldilocks energy landscape (near-perfect) on F-mica, with a dominant energy minima and low barriers to rotation as well as translation, provides a theoretical basis for experimentally-observed epitaxial growth. Further, both CrCl$_3$ and F-mica crystallize in monoclinic lattices, thus promoting aligned growth.

**Patterned Growth and Transfer of Films**

Now we demonstrate proof-of-concept patterned growth using the benefits of substrate-dependent growth (Figs. 5a,b). First, we pattern SiO$_2$ on F-mica substrate (Fig. 5c) via a combination of electron beam lithography (EBL) and electron beam evaporation (Methods). Subsequent growth performed on this patterned substrate (Fig. 5d) shows continuous films on F-mica, terminating at SiO$_2$ regions, illustrating excellent controllability for selected-area 2D-MM synthesis. PL map of patterned growth (Fig. 5e) indicates homogeneous CrCl$_3$ growth on mica, and negligible growth on SiO$_2$ (Supplementary Section S14), reinforcing controllability of selected-area growth, with potential for in-situ device fabrication[39,40].

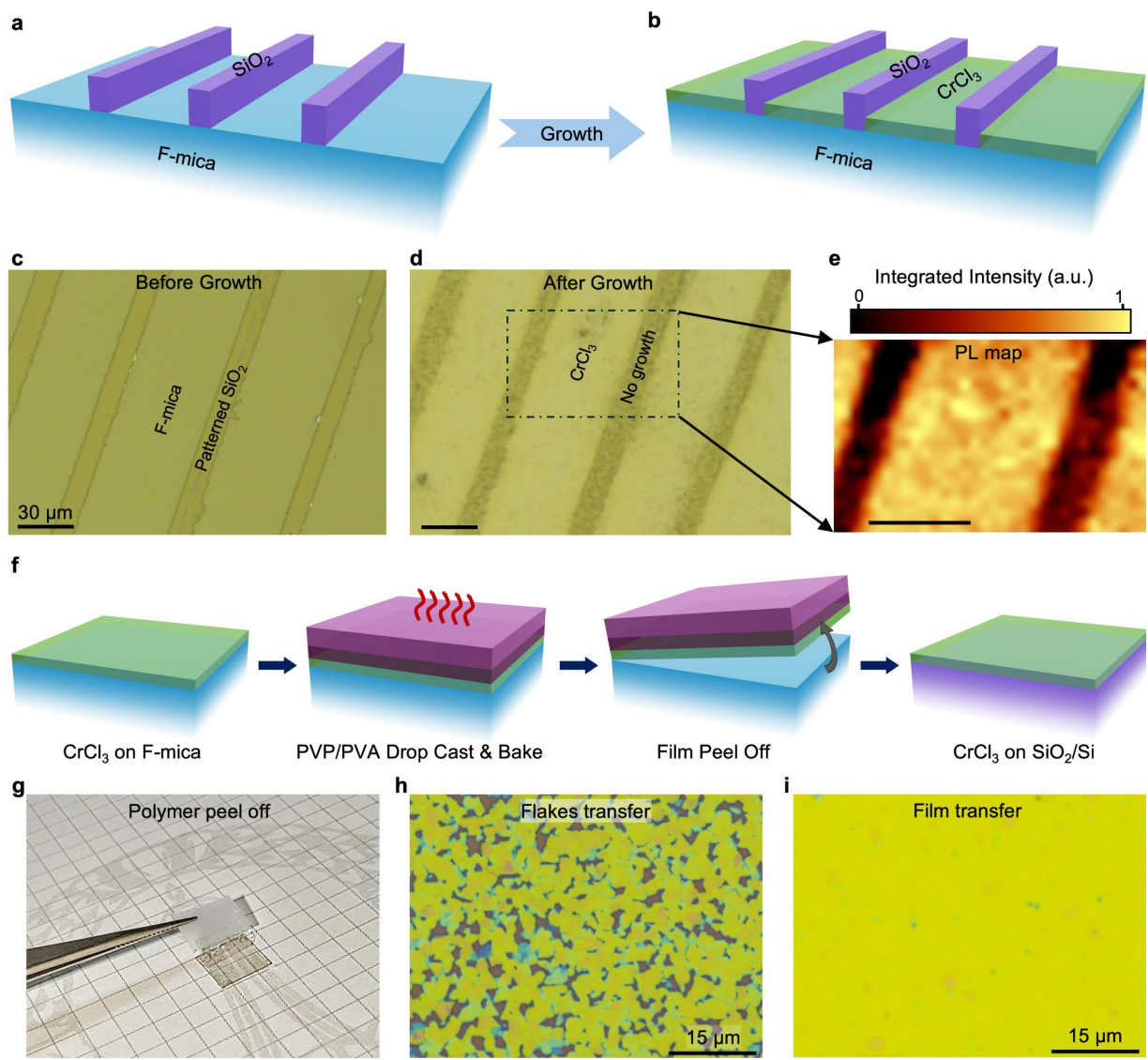

**Fig. 5: Selective-area growth and large-scale transfer. a,b,** Schematic of patterned substrate **a,** before growth, **b,** after growth. **c,** OM image of SiO$_2$ patterned on the F-mica substrate (before growth). The thinner regions are the patterned SiO$_2$. **d,** OM image of the patterned growth shows continuous film growth apart from the patterned SiO$_2$ region. **e,** PL map of the patterned CrCl$_3$ growth marked by the dotted rectangle in **b** confirms continuous CrCl$_3$ coverage except at patterned SiO$_2$ regions. c-e, all scale bars are 30 $\mu$m. **f,** Schematic of steps followed for large-scale transfer. **g,** Camera image taken while peeling off the polymer film. **h,** OM image of flakes and **i,** continuous CrCl$_3$ film transferred on SiO$_2$/Si substrate.

The next critical step towards hetero-integration of 2D-MM is establishing transfer of the grown films. We demonstrate large-scale transfer of CrCl$_3$ from F-mica substrate to target

SiO$_2$/Si substrate. The transfer process (Fig. 5f) involves spin coating PVP (polyvinylpyrrolidone) polymer film and support polymer film, followed by delamination from growth substrate and transfer to target substrate (Supplementary Section S15)[41]. Fig. 5g shows a camera image of the peeling process. Figs. 5h and 5i are OM images of CrCl$_3$ flakes and film respectively, successfully transferred on target SiO$_2$/Si substrate. We note that optimization of adhesion of the PVP polymer is a crucial factor enabling transfer of flakes/films. Transferred films on flexible substrates can be used for strain-dependent magnetization measurements.

**Conclusion and Outlook**

In summary, a tailored PVTD-based approach is developed, leading to wafer-scale, epitaxial growth of 2D-MM CrCl$_3$ on F-mica. Innovative approaches including in-situ light management via low-emissivity secondary heating source, employing oxygen and moisture filters, and precursor flux and temperature control, are critical enablers for synthesizing coalesced films. Substrate temperature was shown to be the control knob for CrCl$_3$ thickness. Optical spectroscopy (Raman, PL), stoichiometric techniques (SEM-EDS, XPS), and TEM confirmed high-quality phase-pure CrCl$_3$. SQUID shows the expected antiferromagnetic-to-ferromagnetic transition around 14 K. Large-scale MLMD simulations unravel the key reasons underlying varying growth on different substrates, providing a significant advancement in theoretical understanding of 2D material growth.

Selective area growth was demonstrated, along with large-area transfer, potentially leading to large-scale devices and straintronics. Since growth is done at low temperatures (~ 500 °C) with a single precursor, there are significant prospects for in-situ integration with other 2D materials for pristine interfaces. For example, CrI$_3$ and CrBr$_3$ powders (and other 2D-MM) are commercially available. PVTD method is inherently flexible for in-situ doping and strain engineering for tuneable magnetic properties. Further, very high-flow rate and low-emissivity secondary sources can be directly applied to other 2D-MM that are oxygen, moisture and light-sensitive. This scalable synthesis approach will lead to incorporation of 2D-MM in spintronic technologies, and for creating heterostructures.

**Methods:**

**Synthesis of CrCl$_3$ film.** PVTD growth of CrCl$_3$ was done in a two-zone furnace (MTI 1200-OTX), with a 1" diameter quartz tube acting as the growth chamber. As shown in Fig. 1a, the CrCl$_3$ powder (Thermo-Fisher, 99.9% (CAS number: 10025-73-7)) and the substrate (sapphire, SiO$_2$/Si, or F-mica) were respectively loaded in 1$^{st}$ and 2$^{nd}$ zone of the chamber. Before the mass flow controller, the inert carrier gas (argon) passes through a two-stage trace moisture and oxygen filter. Further, instead of using regular sealing assemblies (end-caps and O-rings), we use custom KF compression fittings and joined quartz and metal tubes (Supplementary Section 2). These modified fittings lead to much better ultimate vacuum, indicating excellent environmental sealing. Given the atmospheric instability of CrCl$_3$, multiple steps were followed to reduce the chances of degradation of flake during the growth. Firstly, the chamber was evacuated and purged fourfold with ultra-high purity (UHP) argon to remove any residual gases. Then, the chamber was further purged with UHP argon for 30 minutes at room temperature and atmospheric pressure. As an extra precautionary step, under the flow of UHP argon, the furnace was heated to 200°C in 10 minutes and maintained for 60 minutes to remove any moisture from the reactor. Further, the 1$^{st}$ and 2$^{nd}$ zones were respectively heated to 650-725 °C and 450-600 °C in 25 minutes, while the carrier gas flow was maintained. Once heated, the growth was done for 2-30 minutes, during which the carrier gas was flown at a rate of 35-500 SCCM. Supplementary Section 1 demonstrates a typical growth process. Once the growth was completed, the 1$^{st}$ zone temperature was slowly reduced to a lower temperature for one hour, whereas the 2$^{nd}$ zone was continuously naturally cooled.

**Thickness measurement of the flakes.** AFM Park Systems (NX-20) was used to measure the thickness of grown flakes. AFM topography maps were acquired in Non-Contact Mode. Silicon AFM probe from NuNano (Scout 350) with spring constant 42 N/m and resonant frequency of 350 kHz was used for AFM scans.

**Raman and Photoluminescence (PL) characterization.** Raman and PL point spectra were acquired on a HORIBA LabRamHR Raman set-up at room temperature with an excitation wavelength of 532 nm laser and a 50X long working distance microscope objective having a numerical aperture of 0.5. Gratings with 1800 lines/mm and 600 lines/mm were respectively used for Raman and PL spectra acquisition. For all optical measurements, the samples were loaded in a custom-made vacuum chamber having a quartz optical window, and maintained at $\sim 10^{-3}$ mbar.

PL map on patterned growth was recorded on a WITec confocal system (WITec alpha-300R) with a 755 nm CW laser excitation source and a 20X objective lens having a numerical aperture of 0.22, at room temperature, under a vacuum of $\sim 10^{-4}$ mbar.

**SEM and SEM-EDS.** SEM (scanning electron microscopy) imaging was performed in Ultra55 FE-SEM Karl Zeiss at 5 kV accelerating voltage. Stoichiometric analysis of (the elements in) the as-grown flakes was conducted via EDS equipped in the SEM, with point mode performed at 15 kV electron accelerating voltage.

**Transmission electron microscopy (TEM).** For TEM, as-grown flakes were transferred on a PELCO® Silicon Nitride TEM grid (Ted Pella) using a PDMS gel-pack film. The $CrCl_3$ flakes were contacted with PDMS film at room temperature using a custom micro-manipulator[42]. The picked-up flakes were then contacted to the TEM grid at 40 °C, completing the transfer process. The SAED micrograph was acquired in a ThermoFisher® Tecnai™ T20-ST microscope having a LaB6 thermal electron source, at an accelerating voltage of 200 kV. HAADF-STEM imaging was performed in Titan® Themis™ equipped with a monochromator, a CEOS probe corrector for Cs aberration and an X-field emission gun source. For STEM imaging, an acceleration voltage of 300 kV was used, the semi-convergence angle was 24.5 mrad, and the collection angle range of the HAADF detector was 49-200 mrad.

SAED pattern was simulated using the ReciPro software with the zone axis set to [0 0 1], and the electron beam energy set to 200 keV, consistent with experimental conditions[43]. Standard parameters such as camera length and excitation error were adjusted within ReciPro while assuming a dynamic diffraction model. The positions of the diffraction spots were indexed based on the simulated reciprocal lattice of the $CrCl_3$ unit cell. The high-resolution STEM image was simulated using the Dr. Probe software, employing the multislice method with thermal diffuse scattering modelled via the frozen phonon approach[44]. Input parameters such as zone axis, accelerating voltage, convergence semi-angle and detector geometry matched the experimental conditions. The $CrCl_3$ structure (MP-ID: mp-27630) used for simulations was obtained from the Materials Project database, and further, for STEM simulation, the unit cell was orthogonalized using RIPER-Tools[45,46].

**XPS.** XPS measurements were carried out using Thermofisher (Kratos Axis Ultra system) with Al K-α (1.487 keV) x-ray source, 400 μm$^2$ probe, and 50 eV pass energy. The peak fitting and analysis were performed using CasaXPS software[47].

**Magnetic Measurements.** Magnetic measurements were performed using a SQUID magnetometer (MPMS 5XL, Quantum Design). Zero field cooling (ZFC) mode was realized after cooling the sample from 300 K to 2 K in 30 min in zero field. Field cooling (FC) mode was performed with an applied magnetic field (100 Oe) during cooling. Temperature dependence of magnetic moment in a constant magnetic field (200 – 5000 Oe) were recorded during sample heating from 2 K to 30 K, with a heating rate 0.4 K/min (heating mode was "sweep"). Field dependences (hysteresis) were recorded after temperature stabilization at 2 K, 15 K, 16 K, 17 K, and 18 K. The average sweeping rate of the magnetic field was 160 Oe/min.

**Computational methods for the DFT and MLMD simulations.** All quantum-mechanical density functional theory (DFT) calculations were performed using version 5.4 of the Vienna Ab initio Simulation Package (VASP) with spin-polarization turned on and with Grimme's D3 dispersion corrections included with Becke-Johnson damping[48–50]. Dipole corrections were applied to both energies and forces to eliminate spurious interactions along the direction of vacuum[51]. The plane-wave energy cutoff was set to 500 eV, determined through convergence testing of 1 meV/atom accuracy in total DFT energies. The Perdew-Burke-Ernzerhof (PBE) exchange-correlation functional was utilized along with projector augmented wave potentials[52–54]. Hubbard correction ($U = 2.8$ eV) was utilized for the 3d orbitals of Cr[55,56]. The self-consistent field (SCF) convergence criterion was set to 10$^{-6}$ eV. The convergence threshold for geometry optimization using DFT was 10$^{-6}$ eV for energies and 0.01 eV/Å for the maximum magnitude of atomic forces. A 3 × 2 × 1 supercell of F-mica and orthogonal sapphire substrate was utilised for the adsorption energy calculations with at least 18 Å of vacuum.

Machine learning (ML) atomistic simulations were performed using foundation multi-atomic cluster expansion (MACE) models trained on large-scale DFT datasets[20,22]. Two different models were used for initial testing/benchmarking: the MACE OMAT-0 model (9 million parameters), trained on 100 million DFT calculations from the Open Materials 24 (OMAT24) dataset released by Meta, and the MACE MPA-0 model (also 9 million parameters), trained on 12 million DFT calculations from a combination of the MPtrj dataset (from the Materials Project) and a subsampled version of the Alexandria dataset[45,57,58]. The MPtrj and Alexandria datasets primarily consist of near-equilibrium configurations, while the OMAT24 dataset contains a substantial number of non-equilibrium structures obtained from AIMD and rattled DFT structures. Both models were validated and benchmarked against DFT results for key properties such as lattice parameters, adsorption energies, and potential energy curves. Due to its significantly higher fidelity in reproducing DFT results (see Supplementary Section 11), the MACE OMAT-0 model was used for all ML-based results reported in this work. Machine-learning molecular dynamics (MLMD) simulations were conducted using the atomic simulation environment (ASE) python package[59]. These simulations were carried out in the $NVT$ ensemble at 800 K, employing a Langevin thermostat, with a time step of 0.5 fs[60]. A 6 × 4 × 1 supercell was employed for MLMD simulations and rotational PES calculations.

Structural modeling was carried out ASE and RIPER-Tools[46]. Visualizations were produced using CrysX-3D Viewer, while post-processing of MLMD trajectories was done using OVITO package[61,62]. Simulation videos concerning MLMD simulations are available as Supplementary Video Files 1-3.

**Patterning.** The mica substrates were coated with PMMA 495A4. Then, electron beam lithography was performed using Raith Pioneer with an EHT of 20kV and a current of 1.25 nA (60 micrometre aperture). The samples were developed afterwards using 1:3 Methyl isobutyl ketone: Isopropyl Alcohol (MIBK:IPA) solution for 30s, followed by IPA for 30s. 200 nm $SiO_2$ was deposited using Tecport Electron beam evaporator, where $SiO_2$ crystals were evaporated using an electron beam at the rate of 0.5 Å/s. Oxygen was flown continuously during deposition to ensure the $SiO_2$ film was not oxygen deficient. The samples were then kept in acetone for 24 hrs for liftoff. They were further sonicated in acetone for 2 minutes for complete removal of $SiO_2$ from undesired areas. Then, growth was performed as usual.

**Author information:**

Corresponding Author: *Akshay Singh, aksy@iisc.ac.in

**Data Availability**

All data created or analysed during this study are included in this paper and in the Supplementary Information. Additional supporting data are available from the corresponding author upon reasonable request.

**Acknowledgements**

AS acknowledges funding from Indian Institute of Science (IISc) start-up grant and from Department of Science and Technology Nanomission CONCEPT grant (NM/TUE/QM-10/2019). AS, AGR and SP acknowledge funding from Anusandhan National Research Foundation (ANRF) grant SPR/2023/000397. The authors acknowledge the Advanced Facility for Microscopy and Microanalysis (AFMM at IISc), Micro Nano Characterization Facility (MNCF at Centre for Nano Science and Engineering (CeNSE), IISc), National Nanofabrication Center (NNFC at CeNSE, IISc), XPS facility in Department of Inorganic and Physical Chemistry (IISc), and DST-FIST funded Oxford-WITec Alpha 300R Confocal Raman/PL Spectroscope, Central Instruments Facility, Department of Physics (IISc) for use of fabrication and characterization facilities. KSK acknowledges DST-INSPIRE fellowship. MIP and AIC acknowledge the RSCF grant 24-12-00186 for performing magnetic studies. AGR acknowledges the Infosys Foundation, Bengaluru, for the Infosys Young Investigator grant. MS, AGR, SP, and JP thank the Supercomputer Education and Research Centre at IISc for computational facilities. The work was partially supported by the Ministry of Science and Higher Education (FSMG-2025-0005). AS thanks Aveek Bid for providing the two-zone furnace used in the experiments, and Anjana Devi and Harish Parala for discussions on high-vacuum compatible fittings.

**Additional Information**

**Supplementary Information** is available for this paper.

**Correspondence and requests for materials** should be addressed to Akshay Singh

# Supplementary Information

# Tailored Vapor Deposition Unlocks Large-Grain, Wafer-Scale Epitaxial Growth of 2D Magnetic CrCl$_3$

Vivek Kumar[1], Abhishek Jangid[1,#], Manas Sharma[2,#], Manvi Verma[1], Jampala Pasyanthi[2], Keerthana S Kumar[1], Piyush Sharma[2], Emil O. Chiglintsev[3,4], Mikhail I. Panin[3,4], Sudeep N. Punnathanam[2], Alexander I. Chernov[3,4], Ananth Govind Rajan[2], Akshay Singh[1,*]

[1]*Department of Physics, Indian Institute of Science, Bengaluru, Karnataka 560012, India*

[2]*Department of Chemical Engineering, Indian Institute of Science, Bengaluru, Karnataka 560012, India*

[3]*Russian Quantum Center, Skolkovo Innovation City, Moscow 121205, Russia*

[4]*Center for Photonics and 2D Materials, Moscow Institute of Physics and Technology, Dolgoprudny 141700, Russia*

[#]These authors contributed equally.

*Corresponding author: aksy@iisc.ac.in

# Supplementary Information

# Table of Contents

**(1)    Physical vapor transport deposition (PVTD) process for the growth of CrCl$_3$**

Given the environmental sensitivity of CrCl$_3$, several precautionary steps are taken during and prior to growth. Two-stage trace moisture and oxygen traps are used in-line for the inlet gas to further purify the ultra-high purity argon (UHP argon, nominally labelled 99.9999%) used as the process gas. A modified flange connection for improved isolation to the environment is designed and used (Supplementary Section 2). Additionally, an O$_2$ sensor fixed in the line of the growth chamber is used to monitor the O$_2$ level in the system, and growth is usually started when the O$_2$ concentration reaches below 5 ppm.

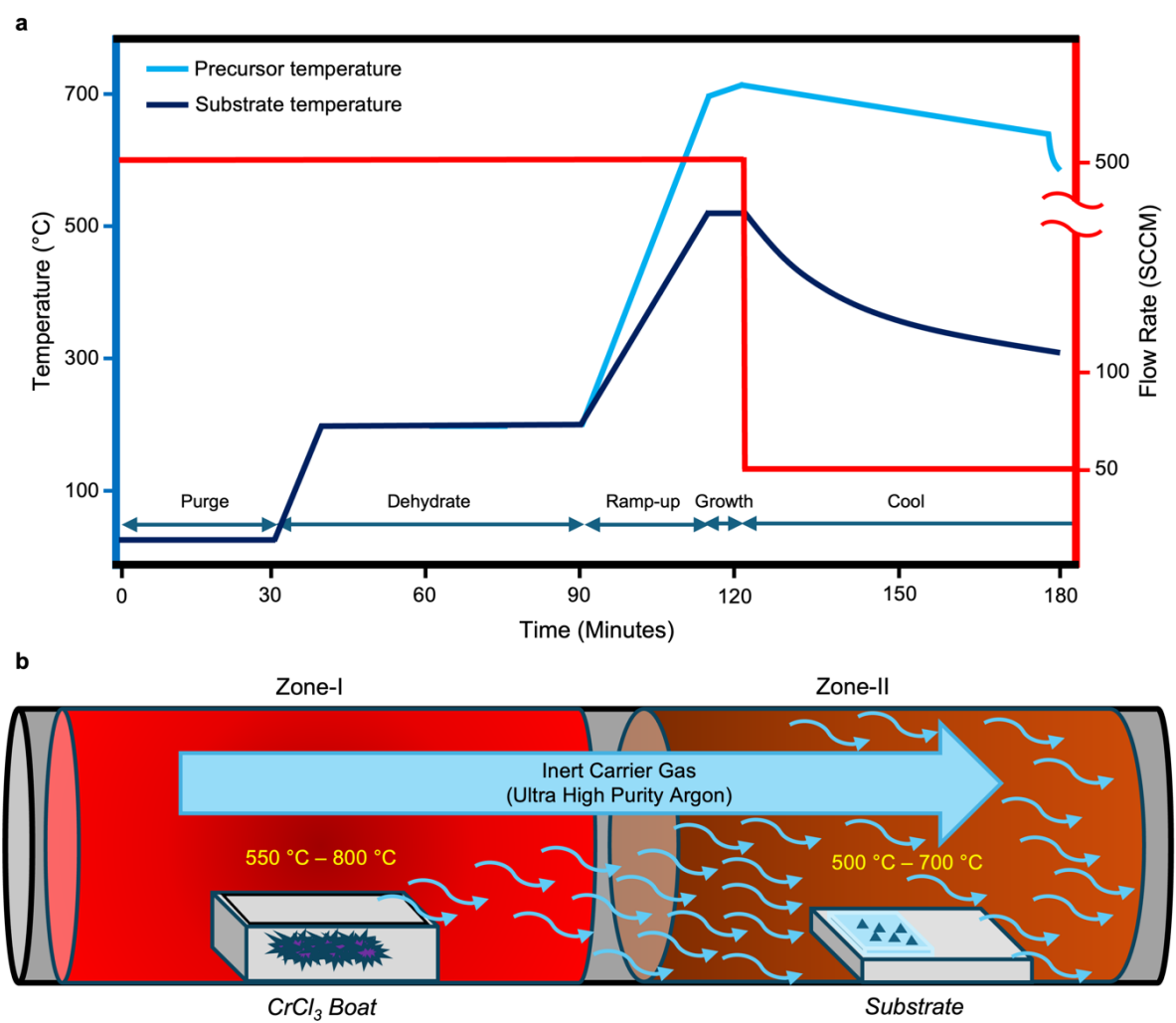

**Fig. S1: Growth of 2D magnetic material (2D-MM) CrCl$_3$. a,** Growth profile for a typical physical vapour transport deposition (PVTD) growth. **b,** Schematic of the growth process.

To mitigate the effect of light-induced damage to the grown films during the growth, where at high temperatures, even a trace quantity of oxygen or moisture could affect the grown films adversely, an aluminium foil is wrapped around the outside of the growth tube to restrict light (also see Supplementary Section 9). Followed by the loading of precursor powder and substrate, the reactor undergoes four evacuation cycles (~1.5 X $10^{-2}$ hPa), and UHP argon is used to vent the system.

Fig. S1a shows a typical growth profile. UHP argon at a flow rate of 500 SCCM is used to purge the system for 30 minutes to remove any remnant gases. To remove any remaining moisture in the system, the reactor is heated to 200 °C from room temperature in 10 minutes, then the system is left for another 60 minutes at 200 °C under the flow of 500 SCCM UHP Ar. The reactor is then ramped to the growth temperature, i.e., for typical growth, zone 1 (precursor) to 675 °C and zone 2 (substrate) to 525 °C in next 25 minutes. Once the growth temperature is achieved, the flow rate is changed to desired value (in given example, 500 SCCM is maintained) during which the substrate temperature is kept constant, and the precursor temperature is ramped further up (700 °C) during the growth (2-30 minutes). Lastly, the flow rate is decreased to 50 SCCM, and the substrate zone is allowed to cool down while the precursor temperature is decreased to 650 °C in 60 minutes. Afterwards, the 1st zone is also allowed to cool to room temperature while the flow rate is still maintained at 50 SCCM. This helps maintain an environment of the precursor during cooling, which can result in higher quality films. Fig. S1b demonstrates the growth process, which includes the evaporation of the $CrCl_3$ precursor in the 1st zone, which is carried to the substrate in the 2nd zone via the inert carrier gas argon, where the growth happens.

**(2)    Flange modification for better insulation to the atmosphere**

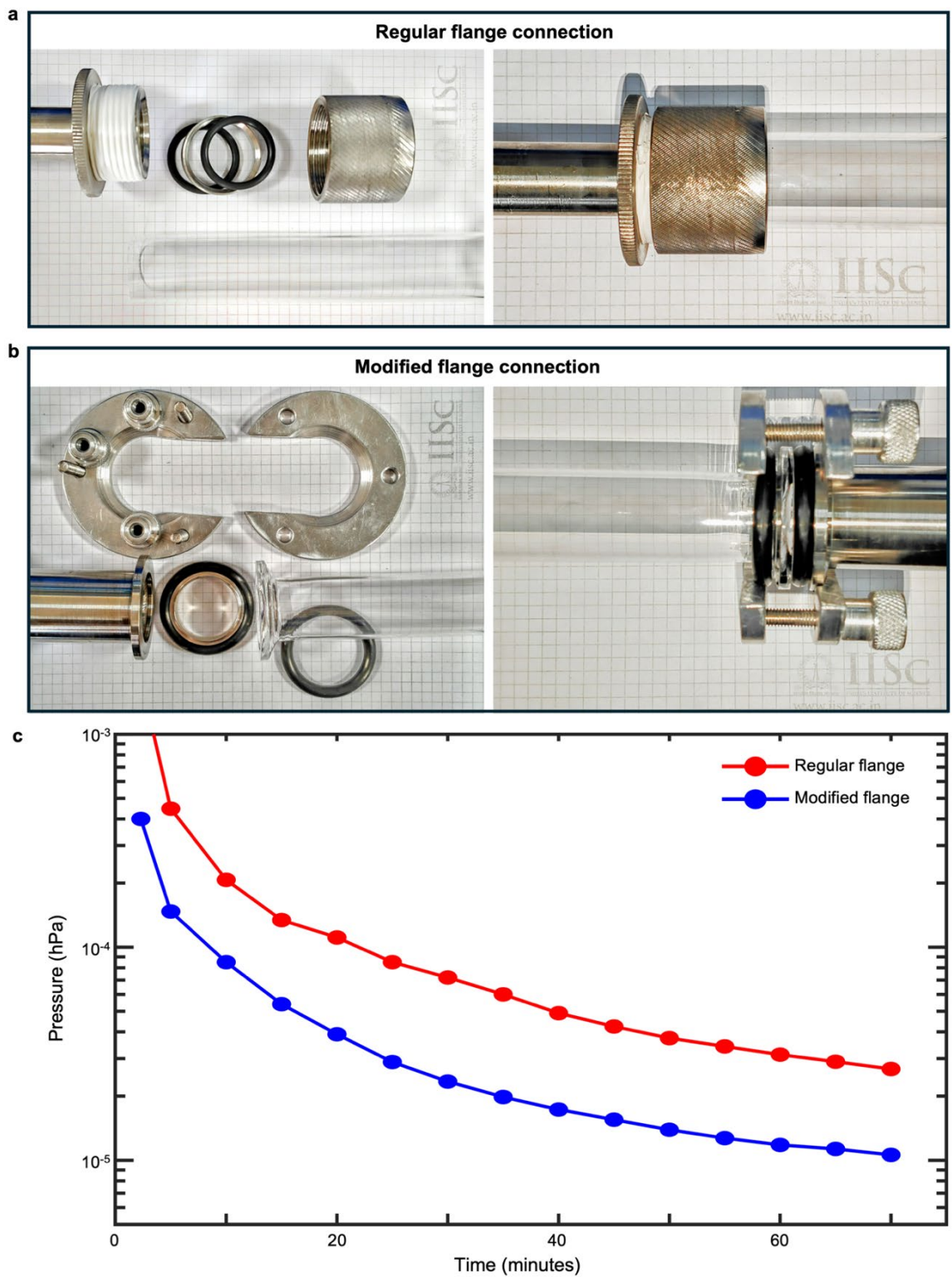

**Fig. S2: Coupling of quartz growth chamber to steel connections.** Components and assembly of **a,** regular and **b,** custom flange connections. **c,** Pressure performance comparison over time shows better atmospheric protection for custom flange connection.

As mentioned, CrCl$_3$ and similar materials are extremely difficult to grow due to their environmental instability in the presence of oxygen or moisture, which is drastically worsened at high temperatures. Hence, reducing every bit of residual oxygen/moisture is of prime importance. To that end, we test the suitability of flange connections, which usually are the primary leak source in reactors. Fig. S2a shows the components (left) and the assembled view (right) of a standard flange connection, which includes threaded fittings and O-rings. Fig. S2b shows a custom flange connection (components (left) and fully assembled configuration (right)), which includes the ends of quartz reactor tube designed as a KF connector and is connected to a steel KF connector using custom clamped fittings. Additionally, a Viton O-ring is used to apply uniform pressure from the KF clamp to the quartz KF connection, which has the added benefit of reduced cracking of quartz tube. Both the connections are tested using the same vacuum assembly, and the modified flange (blue line) consistently achieves and maintains a lower pressure (better ultimate vacuum) compared to the regular flange (red line), indicating superior sealing capability and leak resistance (Fig. S2c). This is the configuration used in our experiments.

**(3)  Large-scale coverage of CrCl$_3$**

The Fluorophlogopite mica (F-mica: (KMg$_3$(Si$_3$Al)O$_{10}$F$_2$)) substrate is imaged after the growth (Fig. S3a) and shows a different contrast compared to the substrates without undergoing growth (Fig. 1b). Additionally, the contrast is consistent and constant over the substrate, depicting the large-scale and uniform growth, further reinforced by the optical microscope (OM) images taken at several positions of the substrate. The substrate is marked by circles at different positions (labelled by B-E), and Figs. S3b-e are the corresponding OM images and show continuous film growth. The large-scale and continuous film growth is further reinforced by the magnified and demagnified OM images (Figs. S3f-i) corresponding to point d.

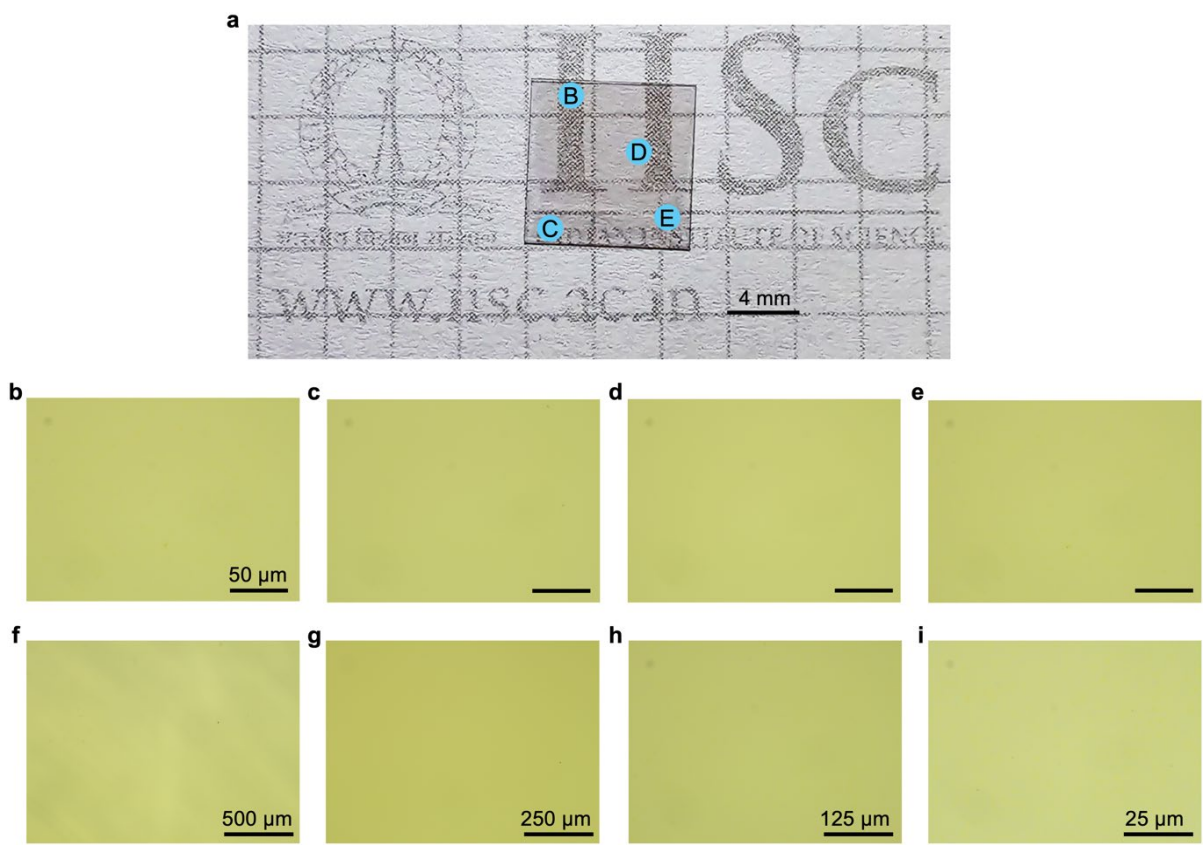

**Fig. S3: Uniform CrCl$_3$ growth over the substrate. a,** Image of the substrate taken from a smartphone after growth shows uniform contrast over the substrate. **b-e,** Respective optical microscope images of the regions of the substrate that are marked by B-E in **a**. Scale bar: 50 μm. **f-i,** Magnified and demagnified OM images corresponding to point D in a.

(4) **Atomic force microscope (AFM) images of as-grown CrCl$_3$ on different substrates**

AFM images of isolated flakes grown on different substrates show clean and uniform topography (Fig. S4a-c). Figs. S4e,f are AFM images of large-scale CrCl$_3$ growth on F- mica substrate, which shows a nearly uniform and continuous film (rms roughness ~ 2 nm), while Fig S4d is AFM image for a growth performed for a shorter time and clearly shows the epitaxial nature of the grown film[1].

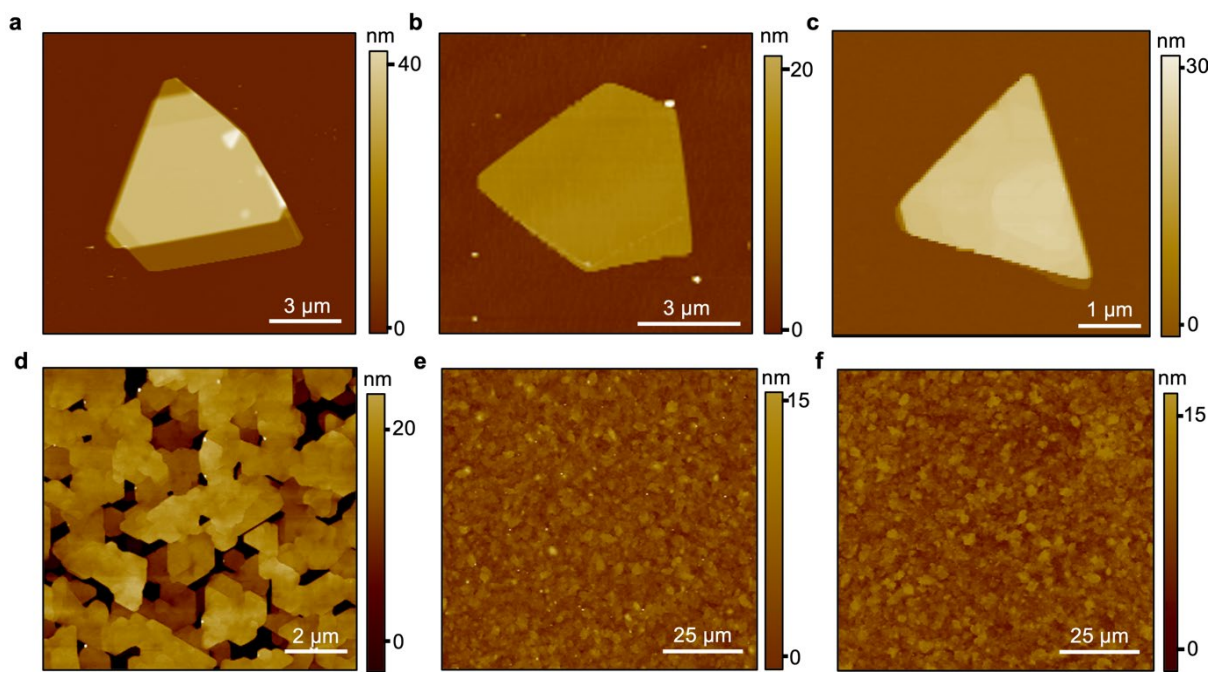

**Fig. S4: Atomic force microscope (AFM) images of the grown CrCl₃ film/flake.** AFM images of isolated thin CrCl$_3$ flakes on **a,** SiO$_2$/Si, **b,** Sapphire (Al$_2$O$_3$), and **c,** Fluorophlogopite mica (F-mica) substrates. **d-f,** AFM images of large-scale CrCl$_3$ film grown on F-mica substrate.

(5)    X-ray photoelectron spectroscopy (XPS) on as-grown CrCl$_3$

XPS analysis of the as-synthesised CrCl$_3$ film on the F-mica substrate confirms high purity. Fig. S5a summarizes the Cl 2p spectrum analysis (from Fig. 3e). It exhibits two dominant spin-orbit components: Cl 2p$_{3/2}$ at 199.19 eV (65.47%) and Cl 2p$_{1/2}$ at 200.80 eV (32.73%) and matches with the peak positions reported in the literature for Cl 2p spectra in CrCl$_3$ and together accounts for 98.2% of the total Cl XPS signal, confirming near-ideal stoichiometry[2]. In addition to the primary peaks, a minor component at 197.95 eV (1.67%) is attributed to a shift due to Cl vacancies. Another weak peak at 200.12 eV (0.13%) corresponds to surface O-(CrCl$_3$) bonding, suggesting minimal (or no) surface oxidation[2]. The overall low defect concentration further supports the high quality and purity of grown CrCl$_3$ films.

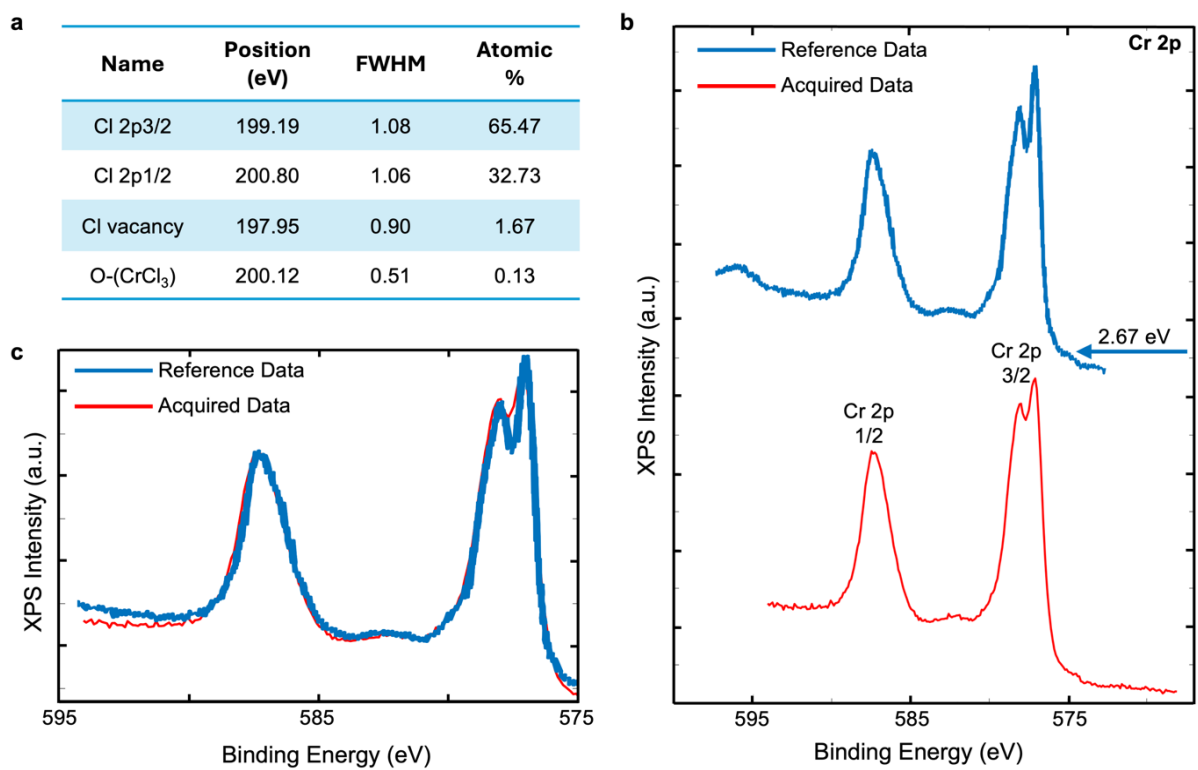

**Fig. S5: X-ray photoelectron spectra (XPS) of grown CrCl₃. a,** Summary of Cl 2p spectra analysis. **b,** Top spectrum (blue) corresponds to the reference data for Cr 2p acquired for CrCl$_3$ and is blue-shifted by 2.67 eV for charge calibration. Bottom spectrum (red) corresponds to acquired Cr 2P of as-grown CrCl$_3$ on F-mica substrate. **c,** Reference[3] and acquired Cr 2p spectra for CrCl$_3$ overlayed on top of each other show an excellent match.

The Cr 2p spectrum (Fig. S5b; bottom) exhibits the characteristic Cr$^{3+}$ multiplet structure, consistent with data for CrCl$_3$ taken from reference (Fig. S5; top)[3]. Due to the inherent complexities of Cr 2p spectra, including multiplet splitting, charge-transfer effects, and shake-up satellites, a full deconvolution is beyond the scope of this work. However, the absence of peaks in the 578–582 eV range, associated with Cr$^{4+}$/Cr$^{6+}$ species, rules out oxidation. Furthermore, the spectral profile closely aligns with that of Cr 2p spectra for CrCl$_3$ in literature (Fig. S5c), reinforcing the conclusion that as-synthesised flakes exhibit extremely high purity[3].
| Name | Position (eV) | FWHM | Atomic % |
|---|---|---|---|
| Cl 2p3/2 | 199.19 | 1.08 | 65.47 |
| Cl 2p1/2 | 200.80 | 1.06 | 32.73 |
| Cl vacancy | 197.95 | 0.90 | 1.67 |
| O-(CrCl$_3$) | 200.12 | 0.51 | 0.13 |

## (6) Transmission electron microscopy of the grown CrCl$_3$ films

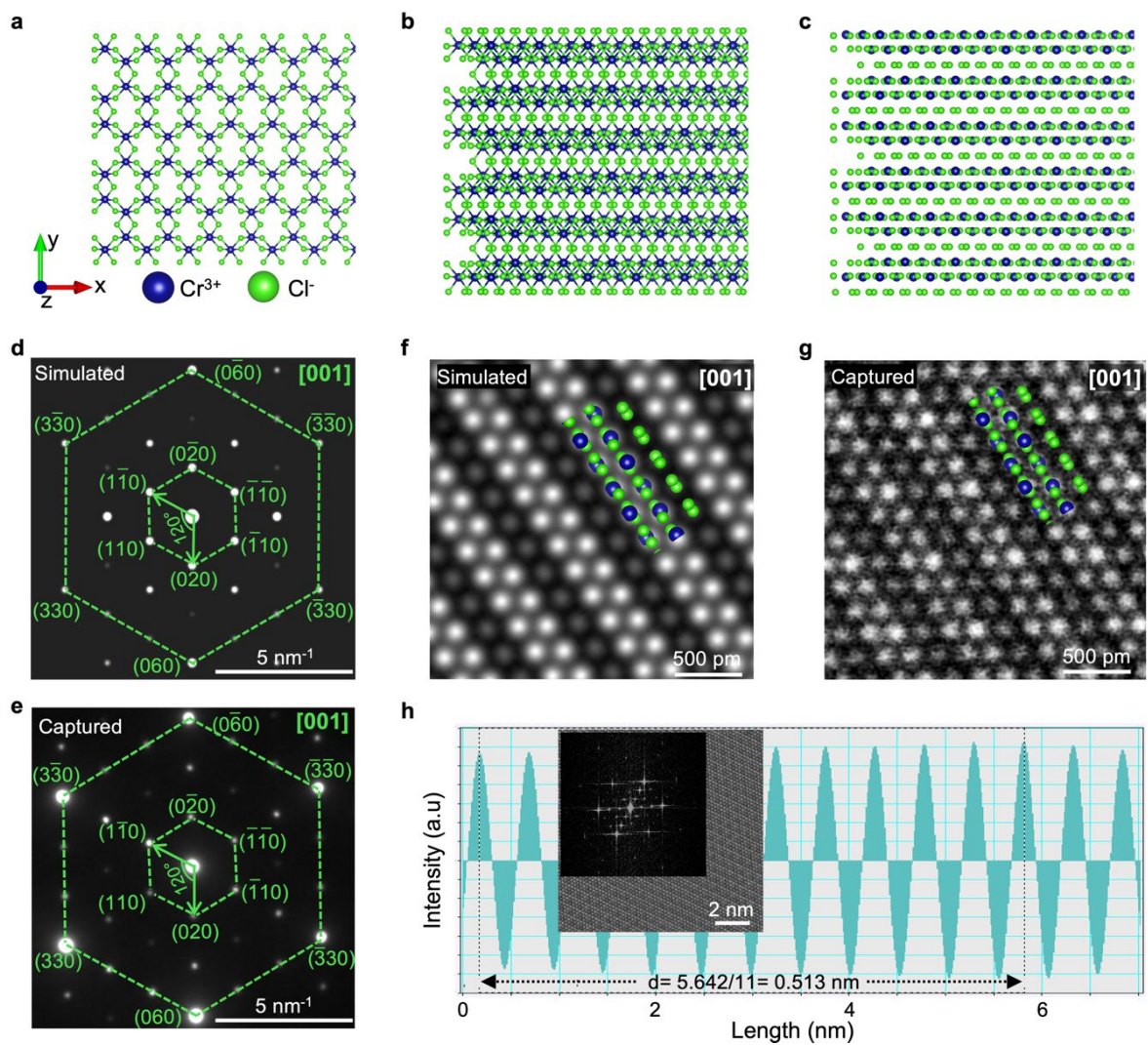

**Fig. S6: Transmission electron microscopy of CrCl$_3$ grown on F-mica substrate. a-c,** Schematic of the atomic structure of monoclinic CrCl$_3$. **a,** Monolayer **b,** Trilayer CrCl$_3$ projected along the Z-axis. **c,** Trilayer CrCl$_3$ without visible bonds. **d,e,** Selected area electron diffraction (SAED) of CrCl$_3$. **d,** Simulated SAED pattern for monoclinic CrCl$_3$. **e,** Captured SAED pattern for CrCl$_3$ matches the pattern in d. **f,g,** Scanning transmission electron (STEM). **f,** Simulated STEM pattern for monoclinic CrCl$_3$. **g,** Captured STEM image of the grown CrCl$_3$ matches the pattern in f. Inset shows an overlay of atomic structure. **h,** d value extracted from Fourier transform of the STEM image in the inset.

CrCl$_3$ exists in the monoclinic phase (space group: C2/m) at room temperature, and the projection of the monolayer along the 001 plane is shown in Fig. S6a, where one Cr atom is connected to six Cl atoms and forms a honeycomb lattice structure[4,5]. Monoclinic $CrCl_3$ follows a ABC-like stacking sequence, and three layers of $CrCl_3$, projected along the z-axis, are shown in Fig. S6b, and without bonds in Fig. S6c for better visualization of the atomic distributions, and show a periodicity that repeats every three atomic rows[6]. Two atomic rows have one Cr atom and two Cl atoms for each column, while the third row has two Cl atoms.

The selected area electron diffraction (SAED) pattern for the monoclinic structure is simulated (Fig. S6d), which matches well with the captured SAED micrograph (Fig. S6e) for the grown $CrCl_3$ sample[7]. The simulated scanning transmission electron microscope (STEM) image (Fig. S6f) matches well with the acquired STEM micrograph (Fig. S6g) for the grown sample[8]. The STEM image exhibits a periodicity of two bright rows of atoms accompanied by a darker row, which matches qualitatively with the observation in Fig. S6c (a portion is overlayed on top), i.e., two rows have two Cl atoms and one Cr atom, while the third row only has two Cl atoms. Additionally, the d-spacing corresponding to 110 planes extracted from the Fourier transform of the low-resolution STEM micrograph is ~ 0.516 nm (Fig. S6h), and matches well with the expected value for monoclinic $CrCl_3$[9].

**(7) SQUID (Superconducting quantum interference device) measurements on the grown $CrCl_3$ films**

Fig. S7a shows the magnetic moment (with diamagnetic contribution of GaAs subtracted) as a function of temperature for out-of-plane fields of 200-5000 Oe for zero-field cooling. Similar to the in-plane configuration (Fig. 2h), the expected sharp kink in magnetization near 14 K is present for out-of-plane magnetic field configurations as well[10].

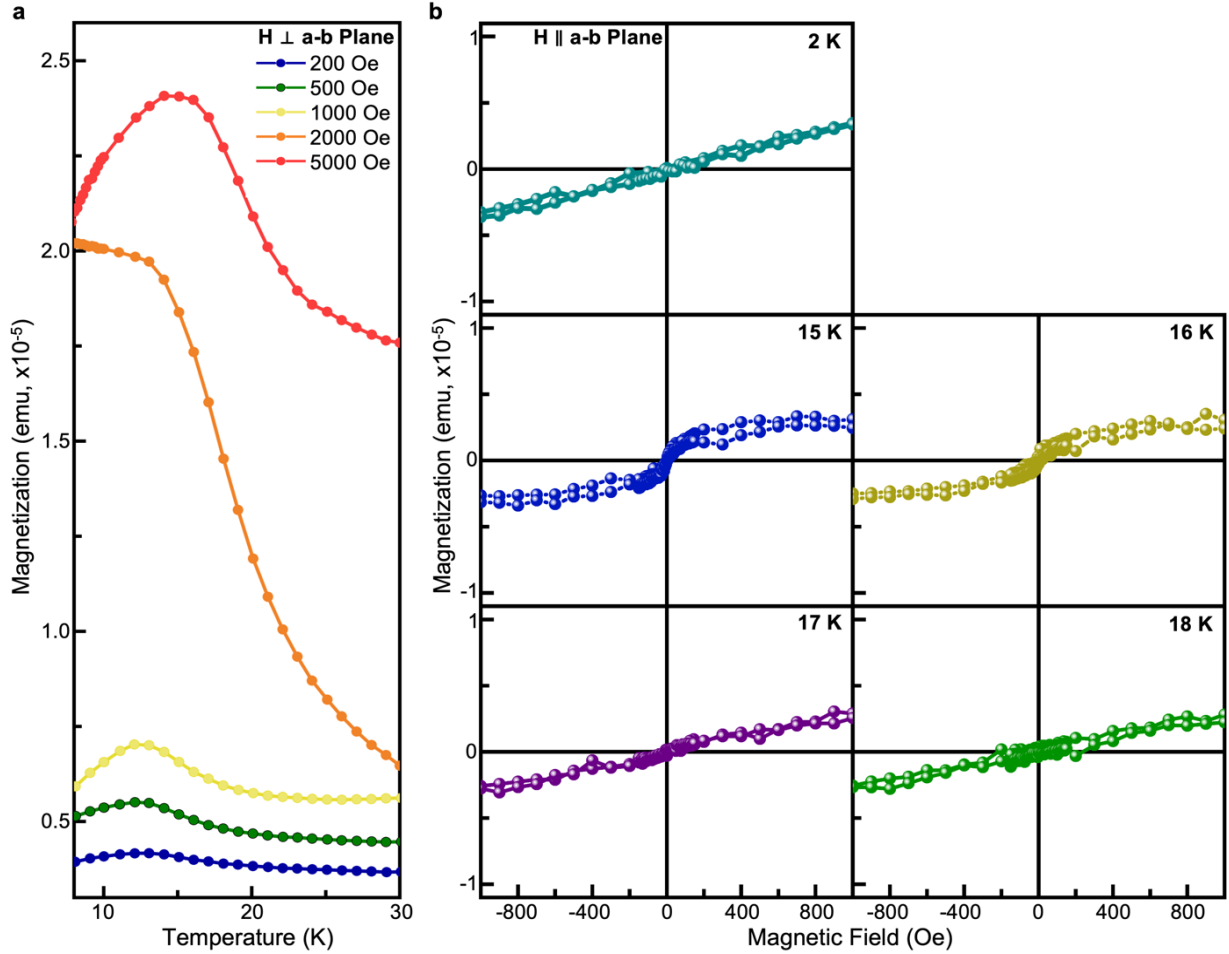

**Fig. S7: Magnetic measurements performed on the CrCl$_3$ films. a,** Magnetization of the sample versus temperature for out-of-plane fields of 200-5000 Oe. The curves for (500, 1000, 2000, and 5000 Oe) are offset in the y-direction, respectively, by (0.64, 1.8, 2.2, and 13.8) × 10$^{-6}$ emu. **b,** Field dependences of the magnetic moment of the sample at different temperatures for in-plane applied field.

To further investigate the antiferromagnetic to ferromagnetic transition around 14 K, the field dependences of the magnetic moment in the range of 2 - 18 K are plotted (Fig. S7b; Contribution of the diamagnetic substrate is subtracted). The magnetic hysteresis is observed in the range of 14-16 K and has the "butterfly" shape (ferromagnetic state). Above 17 K, a linear field dependence is restored (paramagnetic state). This observation confirms the coexistence of the ferromagnetic state at 14-17 K (due to the presence of monoclinic stacking (M-type) in addition to the rhombohedral stacking (R-type)), and a gradual transition to the antiferromagnetic state below 14 K[6]. Additionally, complex combinations of the normal phase with M-type stacking inclusions explain the blurring of the peaks in the temperature dependences at low fields in the perpendicular orientation, as well as the appearance of a plateau at T < 10 K in the parallel orientation. The observed magnetic property matches the reported literature[10].

**(8) Systematic sweep of growth parameters for 2D-MM CrCl$_3$ grown on different substrates**

A systematic parameter sweep for the growth of vdW 2D-MM (CrCl$_3$) on F-mica (Fig. S8i), Al$_2$O$_3$ (Fig. S8ii) and SiO$_2$/Si (Fig. S8iii) substrates is done to understand the effect of each parameter on the growth. While exploring a particular parameter, the rest of the parameters are kept constant. We note that the general trend and explanation for each of the substrates is similar, and hence, we just discuss them in detail for the F-mica substrate.

**i)    Growth on the F-mica substrate:**

The substrate temperature has been identified as a crucial parameter for the growth of 2D-MM. Figs. S8.i.a-d present the evolution of CrCl$_3$ flakes with the increasing substrate temperature from 475 °C to 550 °C while all the other parameters are kept constant, i.e., flow rate is 140 SCCM, growth time is 10 minutes, and precursor (CrCl$_3$) temperature is 688 °C. At 475 °C substrate temperature, small (~2-3 μm) unfaceted thinner flakes are observed. Whereas at 500 °C substrate temperature, larger flakes (~10-12 μm) of triangular and hexagonal shapes of slightly higher thickness are observed. The same trend, i.e., the thickness getting higher, as well as the lateral sizes becoming bigger with the increasing substrate temperature, continues for 525 °C and 550 °C substrate temperatures. In conclusion, we observe that the lateral size and the thickness of the CrCl$_3$ flakes increase with the increasing substrate temperature, and the mechanism is already discussed in the main text.

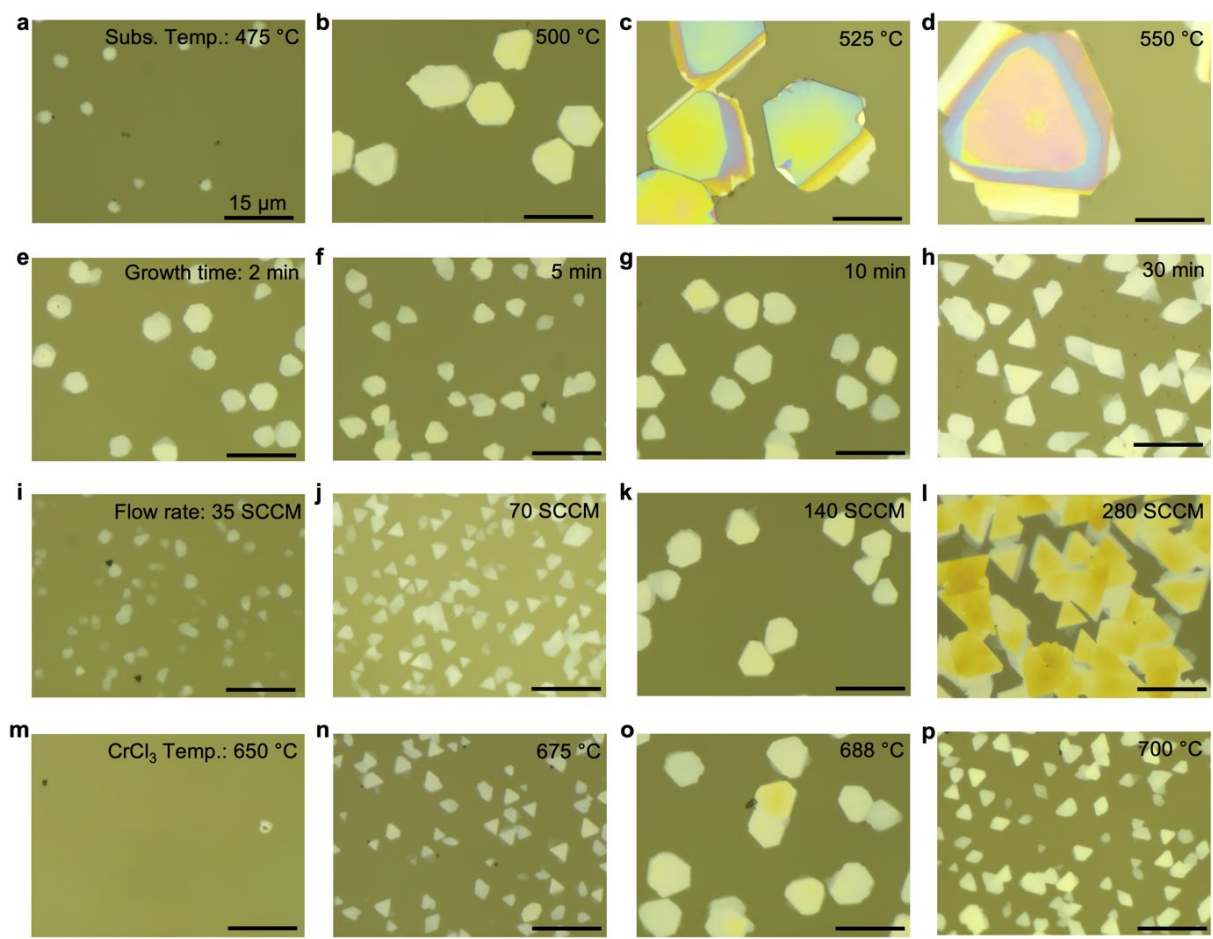

**Fig. S8.i.: Systematic exploration of growth parameters on F-mica substrate. a-d,** Effect of substrate temperature. The lateral size and thickness of the $CrCl_3$ flakes increase with the increasing substrate temperature. **e-h,** Effect of growth time. The size of the grown flakes remains almost constant, but the thickness increases gradually with the increase in growth time. **i-l,** The thickness of flakes increases with increasing flow rate. **m-p,** Effect of $CrCl_3$ temperature on the growth. No growth is observed at a low $CrCl_3$ temperature of 650 °C, while at an increased $CrCl_3$ temperature, growth is observed.

The effect of growth time is mapped in Figs. S8.i.e-h (with the substrate temperature fixed at 500 °C, flow rate at 140 SCCM, and precursor temperature at 688 °C). At a small growth time of 2 minutes, slightly unfaceted thin growth is observed (Fig. S8.i.e). Whereas with increasing time, the flakes become more faceted and thicker (Figs. S8.i.f-h). Interestingly, the lateral size of the grown flakes remains almost constant, whereas the thickness increases with the increase in growth time from 2 minutes to 30 minutes, a characteristic for diffusion-limited growth, attaining saturation early in the process, and has been discussed in the main text[11].

Next, the effect of the carrier gas flow rate during the growth (while the substrate temperature at 500 °C, growth time at 10 minutes, and precursor temperature at 688 °C are kept constant) is explored. Figs. S8.i.i-l are the OM images for the growth done at flow rates of 35 SCCM, 70 SCCM, 140 SCCM, and 280 SCCM, respectively. As expected, the thickness of the flakes increases with the increasing flow rate during the growth due to the increase in precursor supply. Interestingly, the nucleation density (number of flakes per unit area) first increases with the increasing flow rate from 35 SCCM to 70 SCCM, whereas at 140 SCCM, the nucleation density decreases, again increasing at 280 SCCM and can be speculated as - at lower flow rates (35 SCCM, 70 SCCM), although the precursor supply rate is low, the residence time of the precursor at the substrate is large, resulting in higher observed nucleation density. Whereas, at 140 SCCM flow rate, the residence time of the precursor is smaller, explaining the low nucleation density. Lastly, at the flow rate of 280 SCCM, the high precursor supply counteracts the lower residence time and increases the possibility of nucleation.

Lastly, the effect of precursor temperature on the growth (Fig. S8.i.m-p) has been probed at 650 °C, 675 °C, 688 °C, and 700 °C (with substrate temperature fixed at 500 °C, growth time at 10 minutes, and flow rate at 140 SCCM). At a low precursor temperature of 650 °C, an almost blank substrate is observed (Fig. S8.i.m), and almost negligible evaporation of the precursor powder is observed. Interestingly, with just an increase of 25 °C in the precursor temperature, a considerable number of small and thin triangular flakes are observed all over the substrate (Fig. S8.i.n). With a further increase in precursor temperature to 688 °C (Fig. S8.i.o), the sizes of flakes increase to ~ 4 times and can be attributed to the enhanced number of precursor particles over the substrate participating in diffusion-assisted lateral growth. However, with a further increase in the precursor temperature to 700 °C, the sizes of the flakes are observed to decrease, with an increase in number of flakes, likely due to the increased nucleation density at high precursor concentration.

ii)    Growth on the Al$_2$O$_3$ substrate:

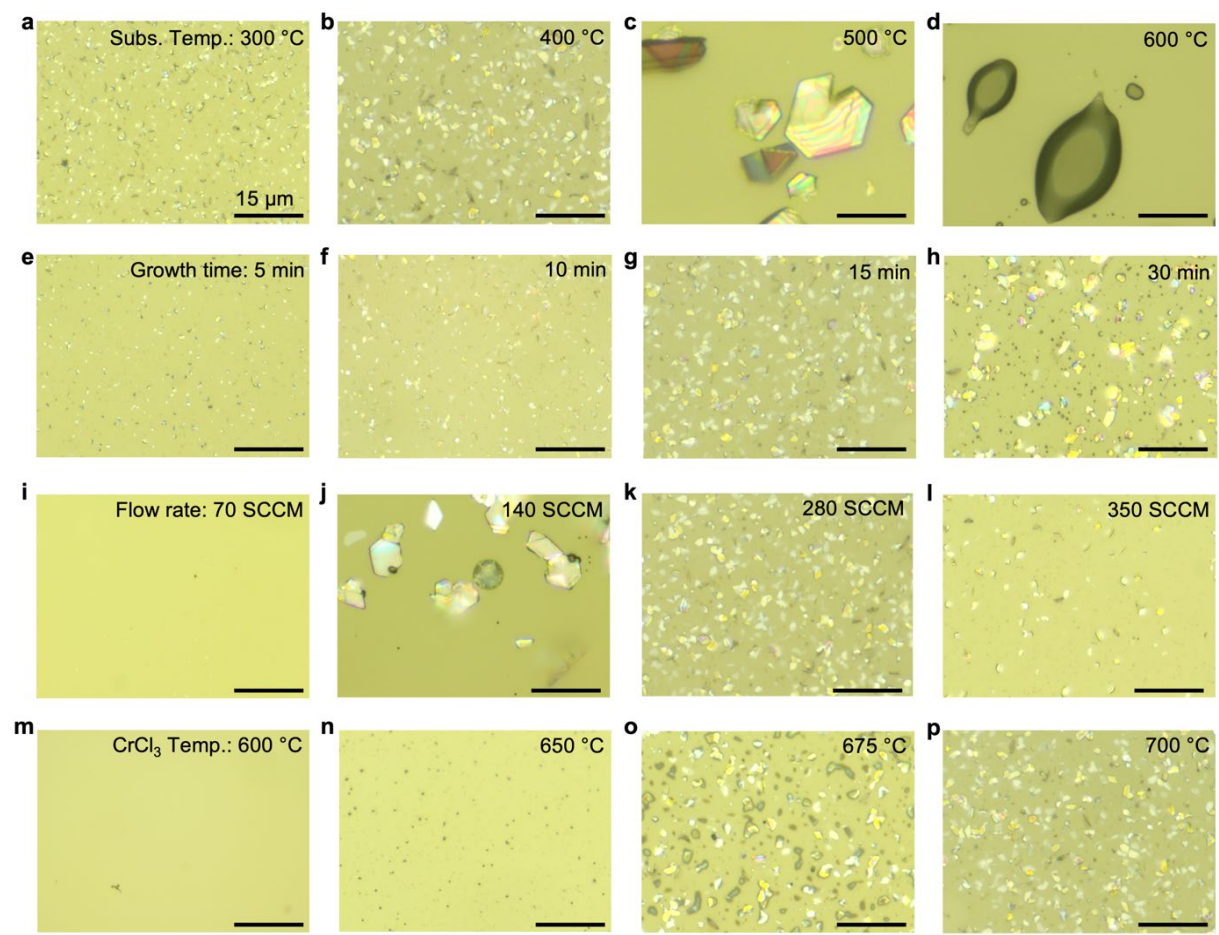

**Fig. S8.ii.: Effect on the CrCl$_3$ growth over Al$_2$O$_3$ substrate with variation in growth parameters. a-d,** The effect of substrate temperature. **e-h,** Optical images of growths with varying growth time **i-l,** The effect of flow rate on the growth **m-l,** The CrCl$_3$ temperature is varied from 600 °C to 700 °C.

**iii) Growth on the SiO₂/Si substrate:**

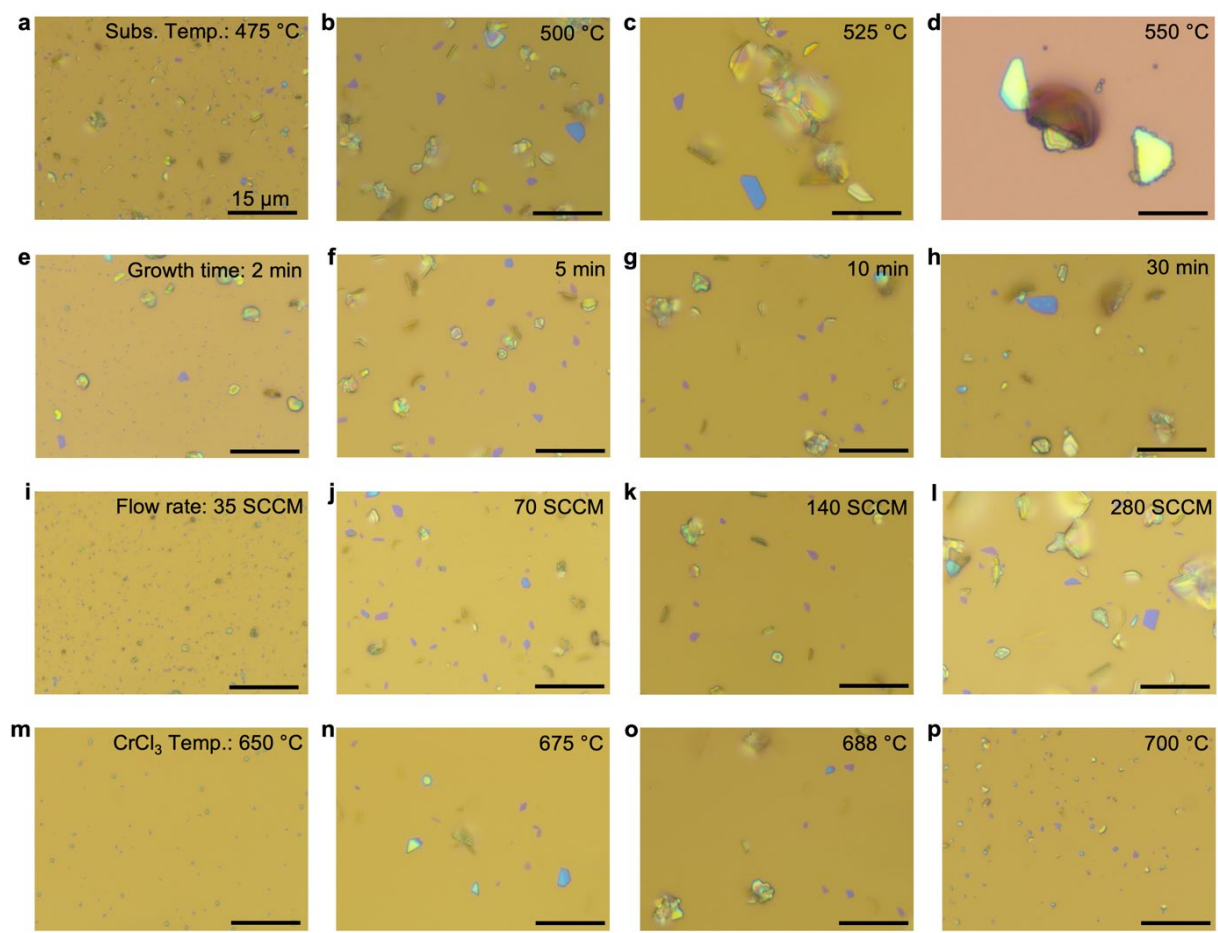

**Fig. S8.iii.: Optimization of parameters for CrCl₃ growth on SiO₂/Si substrate. a-d,** Effect of substrate temperature on the growth. **e-h,** Effect of growth time. **i-l,** Effect of flow rate during the growth. **m-p,** Growth done at different CrCl₃ precursor temperatures.

**(9) Innovations responsible for centimetre-scale growth of 2D-MM**

**Light management:** The heat from the furnace's heating elements within our reactor geometry can be transferred primarily via radiative and conductive components. As mentioned earlier, CrCl₃ and similar materials undergo light-radiation-assisted redox etching in the presence of oxygen and moisture, which is intensified at elevated temperatures. The radiative heat transfer is given by the Stefan-Boltzmann law:

$$Q_{radiation} = \varepsilon\sigma AT^4,$$

where ε is the emissivity of the heating material (range from 0 to 1), and represents a material's ability to radiate energy compared to a perfect black body (emissivity of 1); σ is the Stefan-Boltzmann constant ($5.67 \times 10^{-8}$ W/m$^2$K$^4$); A and T are the surface area and temperature, respectively. Hence, at a fixed temperature and geometry, ε of the source will determine the radiation received by growing CrCl$_3$ and the extent of etching. Typical emissivity value for heating elements used in furnaces is high (0.7-0.88), thus carefully choosing a material with a low emissivity while being a good thermal conductor to act as a secondary heating source is needed. We choose to wrap an aluminium foil, which has a very low emissivity (0.1-0.2) and is a good thermal conductor, around the outside of the quartz reactor tube. An immediate difference is observed in the quality of grown films. Without the aluminium foil wrapping (Fig. 3a, left panel, Main), a severely etched film is produced, whereas with the aluminium foil wrapping (Fig. 3a, right panel, Main), almost no etching is observed.

**Collective innovations:** Reduction in the light received by the growing material, a careful balance between the growth kinetics and thermodynamics, and a continuous increase in the precursor flux during the growth are collectively responsible for the centimetre-scale growth of 2D-MM (summarised in Table S1).

**Table S1: Key growth innovations for substrate-scale growth**

| Al foil wrapping | High flow rate | Precursor ramp during growth |
|---|---|---|
| Light radiation from heating element: ✗ | Precursor chemical potential: ↑ | Precursor flux: ↑ |
| Redox etching: ✗ | Nucleation density and coverage: ↑ | Diffusion-limited growth: ✗ |

## (10) Confirmation of stoichiometric CrCl$_3$ growth on Al$_2$O$_3$ and SiO$_2$/Si substrate

Raman shift spectra of the as-grown flake on Al$_2$O$_3$ and SiO$_2$/Si substrate performed under vacuum are presented in Fig. S9a and S9d, respectively. Six distinct peaks are observed and match the known spectra for monoclinic CrCl$_3$[9,12]. In PL, a broad peak around 1.4 eV, characteristic of CrCl$_3$, is observed for both the as-grown flakes on Al$_2$O$_3$ and SiO$_2$/Si substrate, depicted in Fig. S9b and S9e, respectively[13,14]. Lastly, SEM-EDS presented in Fig. S9c and S9f, respectively, for as-grown flakes on Al$_2$O$_3$ and SiO$_2$/Si, show elemental ratio of Cr to Cl to be nearly 1:3, further confirming stoichiometric growth of CrCl$_3$ on these substrates.

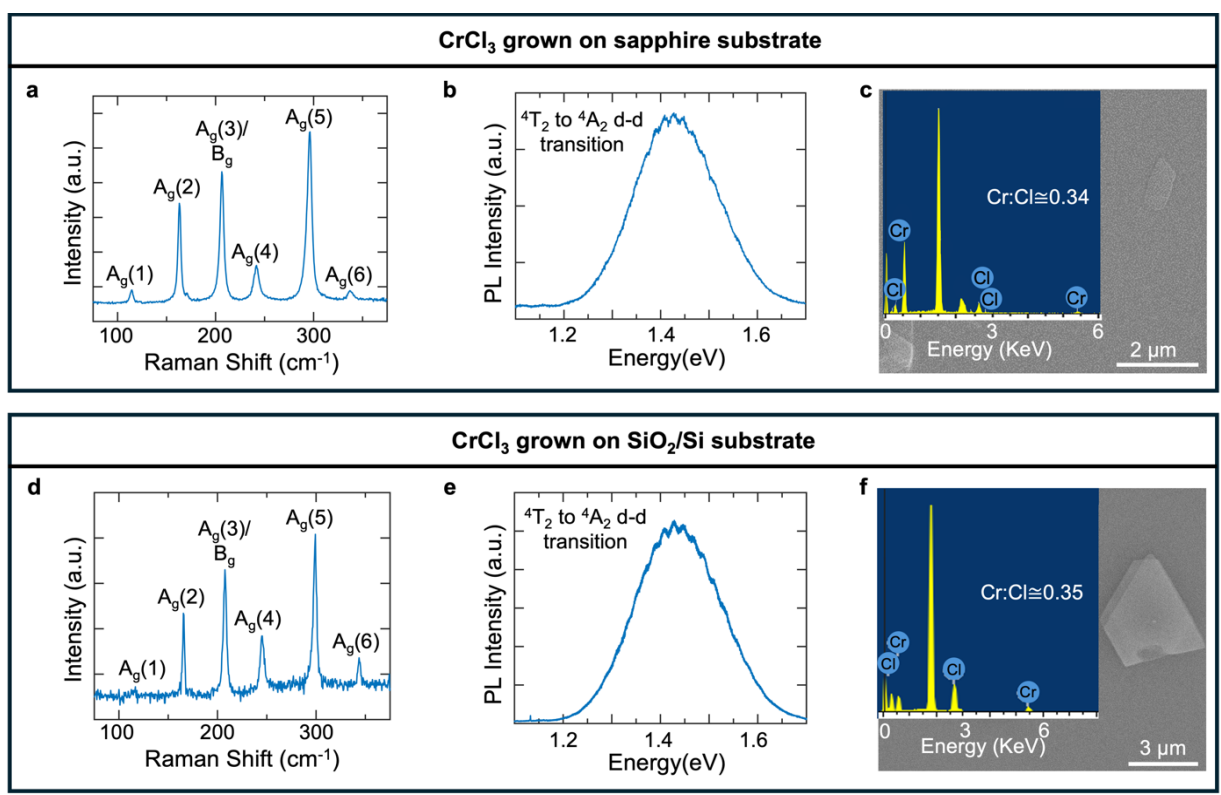

**Fig. S9: Raman, Photoluminescence (PL) spectroscopy and Scanning electron microscope-electron energy dispersive x-ray spectroscopy (SEM-EDS) of CrCl$_3$ grown on Al$_2$O$_3$ and SiO$_2$/Si substrates. a,d,** Raman spectra of CrCl$_3$ flake grown on **a,** Al$_2$O$_3$ and **d,** SiO$_2$/Si matches for monoclinic CrCl$_3$. **b,e,** PL spectra of the as-grown flake on **b,** Al$_2$O$_3$ and **e,** SiO$_2$/Si substrates show the characteristic broad peak (~ 1.4 eV) for CrCl$_3$. **c,f,** SEM-EDS point spectra on the grown flakes on **c,** Al$_2$O$_3$ and **f,** SiO$_2$/Si substrates show near-ideal stoichiometry. Inset shows the corresponding SEM image.

**(11) Validating and benchmarking the machine learning foundation models**

Prior to making predictions with the foundation machine-learned interatomic potentials (MLIPs), a thorough benchmarking and validation process was done to assess the reliability of available MLIP models against established density functional theory (DFT) results. Two prominent foundation models based on the multi-atomic cluster expansion (MACE) architecture were evaluated: MACE MPA-0 and MACE OMAT-0[15,16]. Both models possess ~ 9 million parameters and have been observed to perform quite well in a variety of benchmarks.

The training data composition plays a crucial role in deciding the performance of any MLIP. The MACE MPA-0 model was trained on ~ 12 million DFT calculations from the MPtrj dataset (Materials Project) and a subset of the Alexandria (sAlex) database[17,18]. These datasets primarily contain near-equilibrium structures. In contrast, the MACE OMAT-0 model was trained on a much larger dataset of ~ 100 million DFT calculations from the Open Materials 2024 (OMAT 24) dataset released by Meta (FairChem)[19]. Importantly, the OMAT24 dataset includes a substantial number of non-equilibrium structures derived from *ab initio* molecular dynamics (AIMD) simulations or rattled equilibrium configurations. This suggests that the OMAT-0 MLIP is better poised for the problem at hand, as we perform MLMD simulations which require the model to have a good accuracy for non-equilibrium configurations.

The validation process involves assessing the MLIPs' capability to accurately predict various material properties compared to DFT calculations.

1. **Structural Properties:** We investigate the predictive power of MLIPs in calculating the lattice parameters of sapphire and F-mica bulk structures. Table S2 shows that both MPA-0 and OMAT-0 model predict parameters that are very close to the DFT value and also in good agreement with the experimental results. It is noted that OMAT-0 model consistently performs better than MPA-0, albeit slightly.

| Cell Parameters | Sapphire (Al$_2$O$_3$) | | | | F-mica | | | |
|---|---|---|---|---|---|---|---|---|
| | Experiment[20] | DFT | OMAT-0 | MPA-0 | Experiment[21] | DFT | OMAT-0 | MPA-0 |
| a (Å) | 4.76 | 4.78 | 4.79 | 4.805 | 5.308 | 5.314 | 5.353 | 5.356 |
| b (Å) | 4.76 | 4.78 | 4.79 | 4.805 | 9.183 | 9.228 | 9.288 | 9.291 |
| c (Å) | 12.99 | 13.032 | 13.1 | 13.109 | 10.139 | 10.11 | 10.465 | 10.473 |
| α (°) | 90 | 90 | 90 | 90 | 90 | 89.865 | 89.842 | 89.935 |
| β (°) | 90 | 90 | 90 | 90 | 100.07 | 100.2 | 99.873 | 99.935 |
| γ (°) | 120 | 120 | 120 | 120 | 90 | 90.089 | 90.124 | 90.13 |

Table S2: Unit **cell parameters predicted by DFT and MLIPs (OMAT-0 and MPA-0) for bulk sapphire and F-mica.**

2. **Potential Energy Landscape:** While being able to predict the equilibrium structural parameters accurately is remarkable, it is not enough to deem the MLIPs suitable for MD simulations. Furthermore, MLIPs are to be utilized for investigating the preferred orientation of CrCl$_3$ films on the substrate. All this requires the MLIPs to be able to understand the three-dimensional potential energy surface (PES) of the CrCl$_3$ and substrate system.

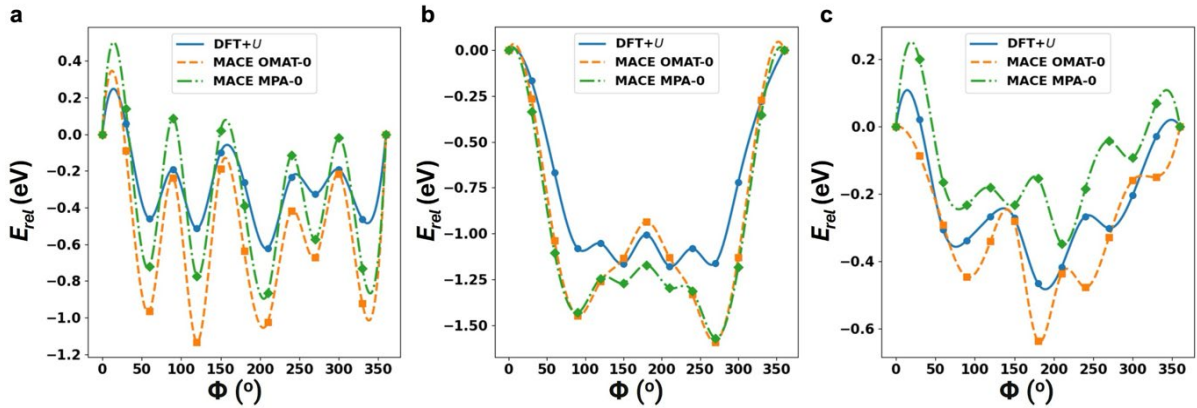

**Fig. S10: Potential energy curves of CrCl$_3$ dimer on various substrates. a-c,** Relative energy, with respect to the initial configuration, as a function of in-plane rotation angle (Φ) of the CrCl$_3$ dimer on **a,** sapphire, **b,** K-terminated mica and **c,** O-terminated mica, with a separation of 3.7 Å between the dimer and the substrate.

To gauge the capability of the MLIPs in capturing the potential energy landscape, we scan the PES of the $CrCl_3$ dimer, in a tilted orientation resembling its configuration in the $CrCl_3$ (001) monolayer, on sapphire and F-mica surfaces (both K and O terminations) for various rotations $\Phi$ (about the surface normal axis) using MLIPs and compare against DFT calculations. This provides a preliminary idea about the MLIPs accuracy in being able to capture the DFT trends.

Both OMAT-0 and MPA-0 reproduce the DFT trends remarkably well, demonstrating strong qualitative and reasonable quantitative agreement. On the sapphire substrate, OMAT-0 better matches the DFT peak positions, while MPA-0 more closely aligns with the valley locations. For the F-mica substrate, OMAT-0 outperforms MPA-0 overall. Despite some deviations, both MLIPs offer reliable qualitative predictions of the DFT PES, highlighting their potential for modelling orientation-dependent interactions in such systems.

3. **Adsorption Energies:** Single $CrCl_3$ molecule adsorption energies on sapphire, F-mica (K-terminated), and F-mica (O-terminated) are calculated using DFT and both MLIPs. The values from DFT, MACE MPA-0 and MACE OMAT-0 models are reported in Table S3. Both MPA-0 and OMAT-0 models predict the trends of strong binding between $CrCl_3$ and sapphire, and weaker binding on F-mica. Interestingly, OMAT-0 performs better for sapphire and MPA-0 performs better for F-mica. Nonetheless, the qualitative agreement of both models with DFT is promising. We also mention in the main text, that we see a confirmation of the strong interaction between $CrCl_3$ and sapphire and weak interaction with F-mica in the MLMD simulations.

| Table S3: Adsorption energy of CrCl₃ on various substrates as predicted by DFT, OMAT-0 and MPA-0. | | | |
|---|---|---|---|
| System | DFT + $U$ (eV) | MACE OMAT-0 (eV) | MACE MPA-0 (eV) |
| CrCl₃ on Sapphire | -5.768 | -3.725 | -1.830 |
| CrCl₃ on Mica (K-terminated) | -0.455 | -0.983 | -0.651 |
| CrCl₃ on Mica (O-terminated) | -0.516 | -0.0002 | -0.033 |

4. **Energies and Forces from AIMD Trajectories**: Short AIMD simulations of 0.25 ps (500 steps with 0.5 fs time step) are performed using DFT for a single CrCl₃ on sapphire and F-mica (K/O terminated) systems. For each frame in these trajectories, we compared the MLIP-predicted energies (relative to the initial frame) and forces with the corresponding DFT values. These comparisons, shown in the parity plots in Figure S11, provide a realistic assessment of how well the MLIPs capture the potential energy landscape during dynamic simulations.

Both MPA-0 and OMAT-0 perform impressively. However, the OMAT-0 MLIP results in higher $R^2$ and lower errors for all cases except relative energies of CrCl₃ on sapphire and forces of CrCl₃ on F-mica (O-terminated). For sapphire, OMAT-0 and MPA-0 yield a mean absolute error (MAE) of 3.5 meV/atom and 3.3 meV/atom for relative energies, respectively, and MAE of 64.3 meV/Å and 84.7 meV/Å for forces. For K-terminated mica, the MAE for relative energies is 3.6 meV/atom using OMAT-0 and 4.7 meV/atom using MPA-0. The MAE for forces for K-terminated mica is 62.0 meV/Å and 74.6 meV/Å using OMAT-0 and MPA-0, respectively. The errors are slightly larger for O-terminated mica, which is somewhat expected, as K-terminated mica is known to be the natural configuration and also similar to the bulk structure.

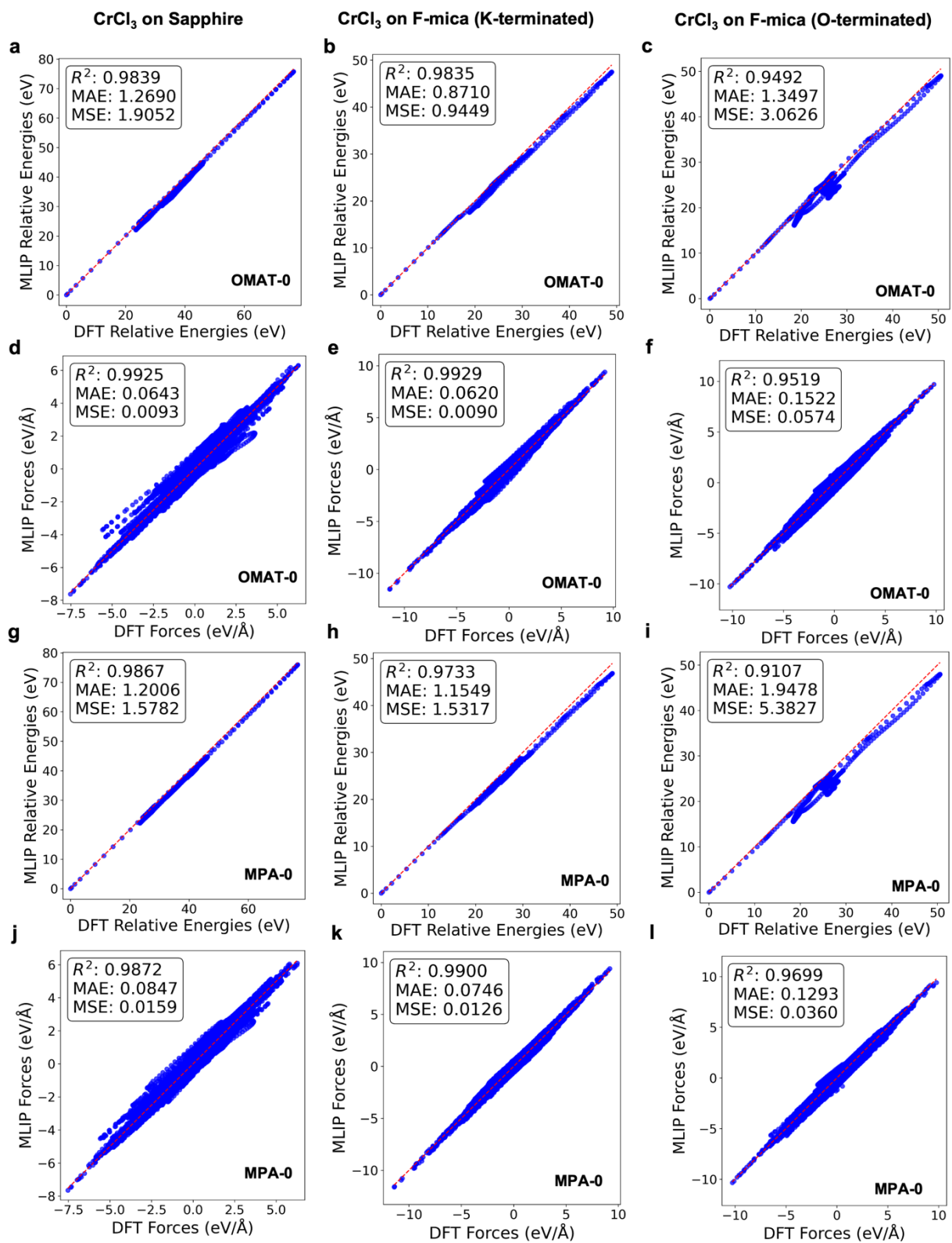

**Fig. S11: Parity plots comparing DFT reference data and MLIP predictions. a–c,** Parity plots for total energies relative to the first AIMD frame (in eV) predicted by the OMAT-0 model for $CrCl_3$ on different substrates: **a,** sapphire, **b,** F-mica (K-terminated), and **c,** F-mica (O-terminated), respectively.

**d–f,** Corresponding parity plots for atomic forces (in eV/Å) predicted by the OMAT-0 model for the same substrates. **g–i,** Parity plots for relative energies predicted by the MPA-0 model for CrCl$_3$ on **g,** sapphire, **h,** K-terminated F-mica, and **i,** O-terminated F-mica, respectively. **j–l,** Corresponding force parity plots for the MPA-0 model across the same substrates. In each panel, DFT reference values are plotted along the *x*-axis and MLIP predictions along the *y*-axis. The red dashed line indicates perfect agreement ($y = x$). Each plot also includes the coefficient of determination ($R^2$), mean absolute error (MAE), and mean squared error (MSE) to quantify prediction accuracy.

Interestingly, both OMAT-0 and MPA-0 perform well in the benchmarks shown above, with OMAT-0 exhibiting a slight edge based on the parity plots. While either model could, in principle, be used to investigate the present systems, the MACE OMAT-0 model was ultimately chosen for all subsequent ML-based simulations due to its broader training set, which includes a larger number of non-equilibrium structures.

### (12) MLMD of CrCl$_3$ molecules on different substrates

Here, we present additional snapshots and analyses of the MLMD trajectories for 20 CrCl$_3$ molecules on sapphire and F-mica substrates, as discussed in the main text. Figure S12 displays top and side views of the atomic displacement vectors after 40 ps of MLMD simulation in the *NVT* ensemble at 800 K. Lower panel in Figure S12 shows the trajectory paths of all atoms over the same time span. The displacement vectors and trajectory lines reveal limited movement of CrCl$_3$ molecules and clusters on the sapphire surface, in contrast to significantly greater mobility on both K-terminated and O-terminated F-mica. Notably, side-view snapshots indicate that the CrCl$_3$ species remain at a relatively large vertical distance from the O-terminated mica surface, suggesting a lack of binding. This observation is consistent with the low adsorption energies reported in Table S3.

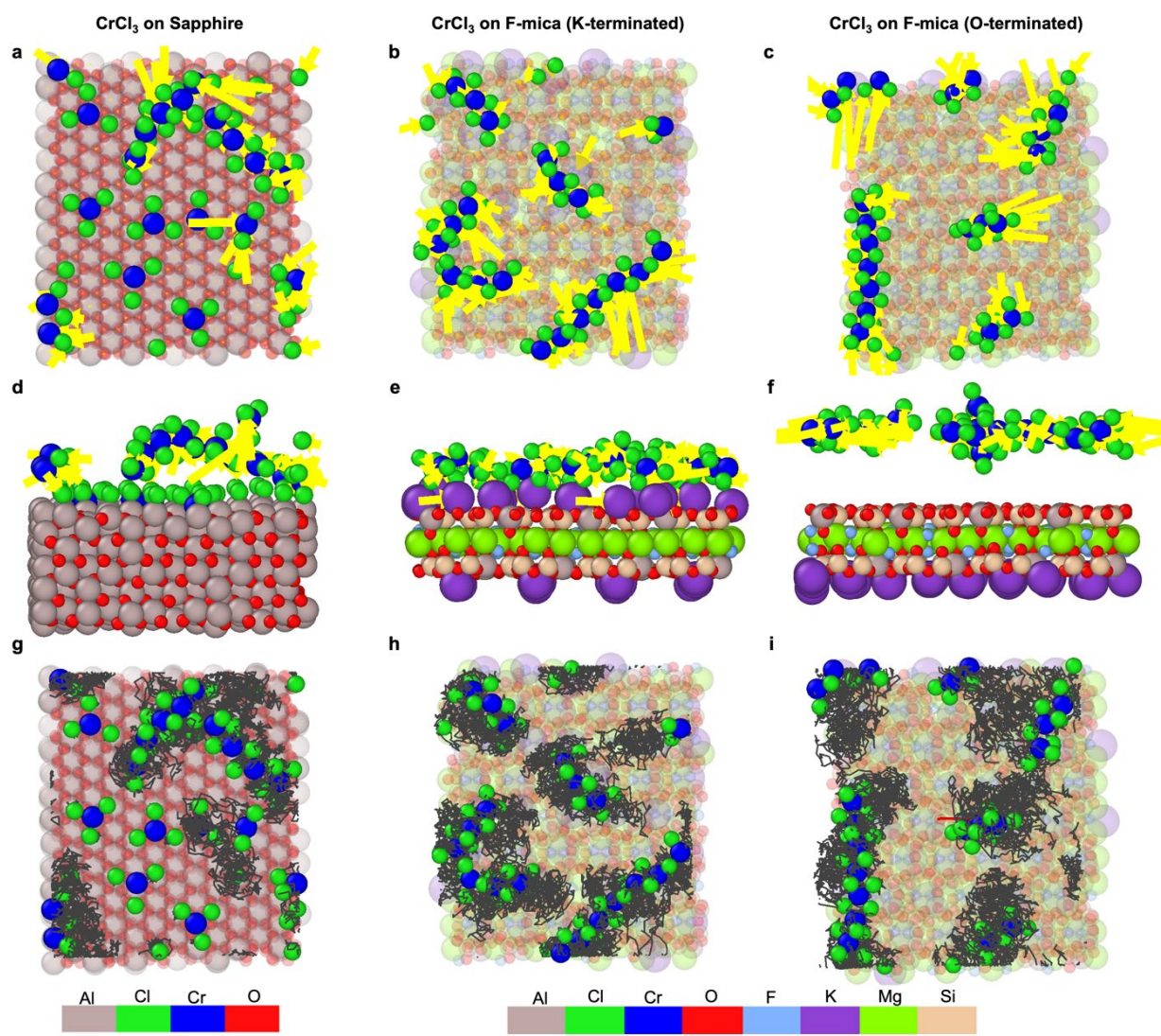

**Fig. S12: Displacement vectors and atomic trajectories of CrCl$_3$ molecules on various substrates from MLMD simulations. a–c,** Top views of atomic displacement vectors over a 40 ps MLMD trajectory (*NVT* ensemble at 800 K) for 20 CrCl$_3$ molecules adsorbed on **a,** sapphire, **b,** K-terminated F-mica, and **c,** O-terminated F-mica substrates. Yellow arrows represent the net displacement of each atom between the start and end of the simulation. **d-f,** Corresponding side views of the same configurations shown in panels **(a–c),** illustrating the vertical positioning and motion of CrCl$_3$ molecules relative to the substrate. **g–i,** Top-down projections of the full atomic trajectories traced over the 40 ps simulation for the same systems: **g,** sapphire, **h,** K-terminated F-mica, and **i,** O-terminated F-mica. These trajectory lines (in gray) show the paths traveled by atoms throughout the simulation.

**(13) Potential energy curve for translating a CrCl₃ molecule**

The adsorption energy of a CrCl$_3$ molecule on sapphire is an order of magnitude higher than that on F-mica (for both K- and O-terminations) as seen in Table S3, suggesting restricted mobility on the sapphire surface. To further explore this behaviour, we calculate the PES for lateral translation of a CrCl$_3$ molecule along the *x*-direction on both sapphire and F-mica substrates using the OMAT-0 MLIP. The initial and final configurations are fully relaxed with OMAT-0 MLIP, and intermediate configurations are generated through linear interpolation. Single-point energy calculations are then performed to construct the PES. Only the K-terminated F-mica surface is considered, as the OMAT-0 MLIP predicts negligible binding on the O-terminated surface (see Table S3), resulting in a flat PES. The resulting PES (Figure S13) reveals a substantial translational barrier of 10.033 eV on sapphire, indicating strong localization, whereas the barrier on K-terminated F-mica is much lower (0.054 eV), allowing for easy molecular movement across the surface.

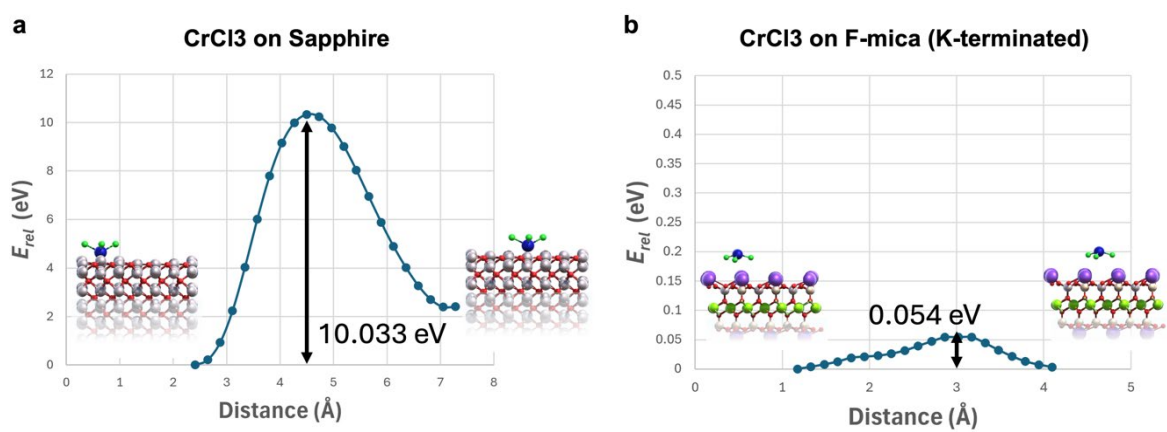

**Fig. S13: Translational energy barrier of a CrCl$_3$ molecule. a–b,** Potential energy surface (relative to the initial frame) as a function of distance travelled by the CrCl$_3$ molecule in the *x*-direction on **a,** sapphire and **b,** F-mica substrates.

Finally, we note that literature suggests the natural configuration of F-mica is K-termination. Only under specific treatment of the substrate, O-termination becomes stable.

Hence we have performed calculations for both K- and O-terminations, but show K-termination in the main text[22]. Importantly, we observe a close agreement of experimental data and K-termination calculations, justifying the focus on K-termination.

**(14)   Selected area growth of CrCl$_3$**

The nature of growth on different substrates is different (Fig. 4, Main), which enables us to demonstrate the proof-of-concept patterned growth of 2D-MM (Fig. 5, Main). Firstly, a freshly cleaved F-mica substrate is spin-coated with polymethyl methacrylate (PMMA) polymer, which is then selectively (in straight lines) irradiated by electrons (electron-beam lithography), followed by MIBK/IPA development of the PMMA, which removes PMMA from the electron-illuminated areas. 200 nm of SiO$_2$, using an electron beam, is then deposited on the substrate, in the presence of oxygen, which is followed by the liftoff of PMMA/SiO$_2$ in acetone, resulting in SiO$_2$ lines patterned on the F-mica substrate (Fig. S14a). Regular PVTD growth of CrCl$_3$ is then done on the patterned substrate (Fig. S14b), showing continuous CrCl$_3$ films on the F-mica region, which is terminated at the SiO$_2$ regions, demonstrating a superb control of selective-area synthesis 2D-MM. PL mapping (step size of 3 μm, integration time of 2 seconds) is performed with a 755 nm continuous-wave laser (in vacuum) to further demonstrate the excellent controllability of the selected-area growth. Fig. S14c is the PL map (intensity integrated for emission energy of 1.35-1.55 eV) overlayed on the top of corresponding OM images, which shows homogeneous growth of CrCl$_3$ on the F-mica region, and a negligible growth on the SiO$_2$ regions, reinforcing the potential for in-situ device fabrication for 2D-MM[23,24].

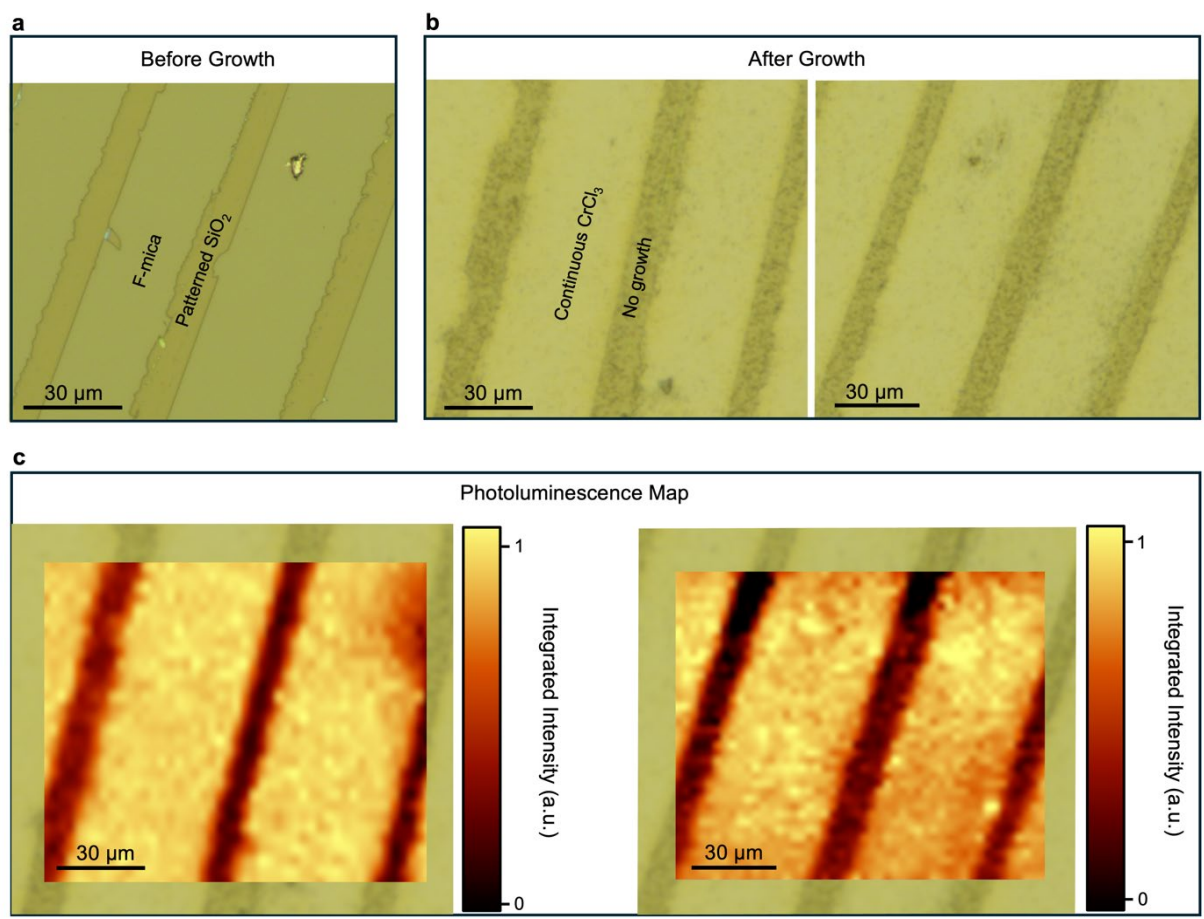

**Fig. S14: Patterned Growth. a,** SiO$_2$ patterned F-mica substrate. **b,** Large-area continuous CrCl$_3$ film grown selectively on the F-mica region. **c,** PL map corresponding to the CrCl$_3$ peak of the patterned growth overlayed on the corresponding OM image.

### (15) Large-scale transfer

For the large-scale transfer of CrCl$_3$ from the as-grown substrate, firstly, 1.5 g of Poly-Vinyl Pyrrolidone (PVP, Alfa Aesar, average M.W. 58000), 1.5 ml of N-Vinyl Pyrrolidone (NVP, Sigma-Aldrich, 99%) were mixed in ethanol to obtain a final solution volume of 10 ml at room temperature[25]. This solution was spin-coated onto as-grown CrCl$_3$ on F-mica substrate at 3000 RPM for 1 minute, followed by baking at 70 °C for 1 minute. Then, to act as a support layer, a 9 wt.% aqueous solution of Poly-Vinyl Alcohol (PVA) was prepared and spin-coated twice onto the PVP-coated substrate, firstly at 3000 RPM for 1 minute and then at 500 RPM for 30 seconds. The substrate was baked at 70 °C for 1 minute after each PVA coating, resulting in the evaporation of solvent and the formation of a rigid polymer film.

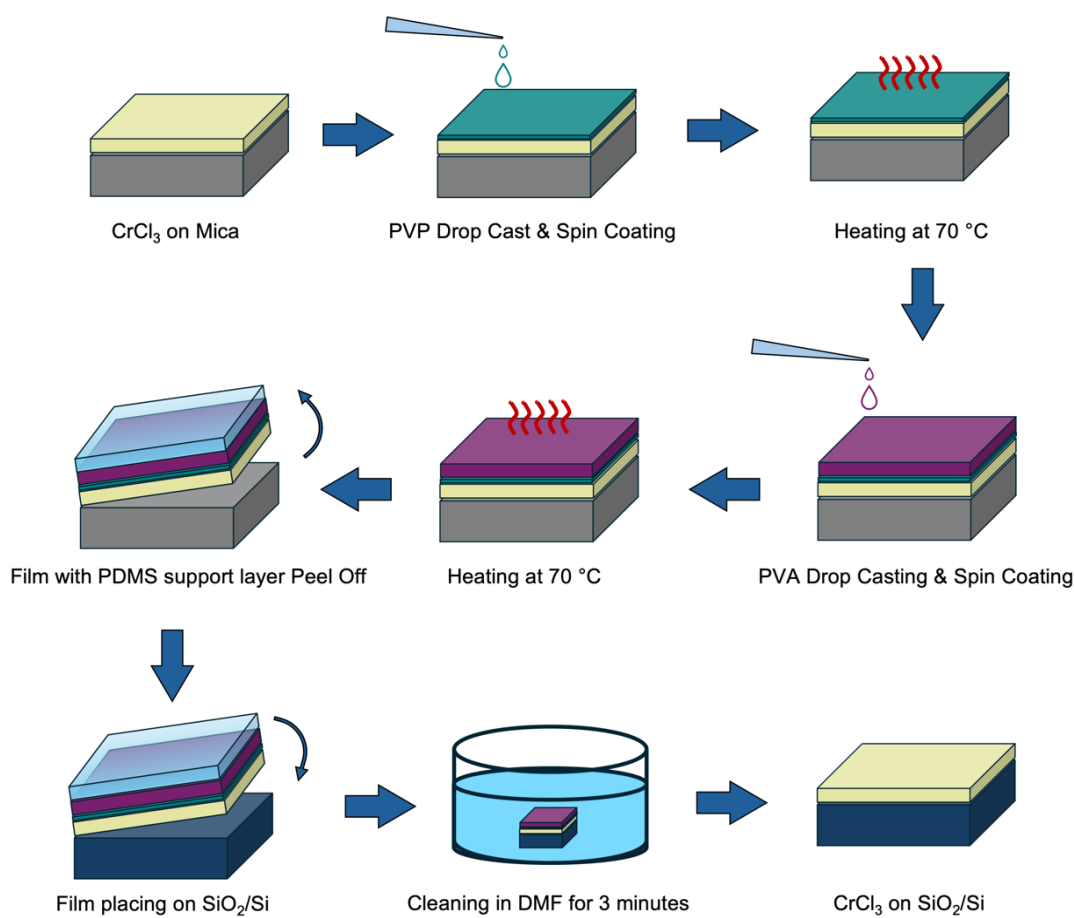

**Fig. S15: Steps followed for the large-scale transfer of as-grown CrCl$_3$ film.**

An edge of the polymer-coated substrate was then scratched using a surgical knife to ease the delamination process of the polymer film. A piece of Scotch tape was stuck to the scratched edge to aid in lifting the polymer, and PDMS film (Gel Pak) of dimensions equal to the substrate was cut and positioned on top of the polymer film to provide further support. The assembly was gently peeled off from the scratched corner. During this process, CrCl$_3$ film was transferred to PVP/PVA film from F-mica substrate due to high adhesion between CrCl$_3$ and PVP compared to CrCl$_3$ and F-mica.

The polymer film assembly consisting of CrCl$_3$ film was then carefully transferred to a SiO$_2$/Si target substrate, followed by gentle application of uniform pressure on the polymer on top of the substrate, which ensured conformal contact between the flake/film and the substrate. PDMS film was carefully removed with the help of tweezers, leaving the polymer film on the SiO$_2$/Si substrate. The assembly was then immersed in N,N-Dimethylformamide (DMF) at 140 °C for 3-5 minutes to dissolve the polymer film. After the removal of the polymer film, CrCl$_3$ film remains adhered to SiO2/Si substrates, thus marking the successful transfer (Main text Fig. 5g-i).